\documentclass[useAMS,usenatbib,twocolumn,fleqn]{mn2e}
\usepackage{natbib, graphics, epsfig}
\usepackage{color}

\usepackage{times}
\usepackage{amsmath}
\usepackage{amssymb}
\usepackage{textcomp}
\usepackage{verbatim}
\usepackage{threeparttable}

\newcommand{\msun}{{\,\rm M_\odot}}

\newcommand{\Gyr}{\,{\rm Gyr}}

\newcommand{\pc}{\,{\rm pc}}

\newcommand{\kpc}{\,{\rm kpc}}

\newcommand{\hmpc}{\,h^{-1}\,{\rm Mpc}}

\newcommand{\ion}[2]{\hbox{#1\,{\sc #2}}}

\newcommand{\HI}{\ion{H}{i}}

\def\jcap{J. Cosmol.  Astropart. Phys.}
\def\aap{A\&A}
\def\apj{ApJ}

\def\apjl{ApJ}
\def\mnras{MNRAS}

\def\aj{AJ}

\def\nat{Nature}

\def\prd{Phys. Rev. D}
\title[Dwarf galaxies in CDM and SIDM with baryons]
      {Dwarf galaxies in CDM and SIDM with baryons: observational probes of the nature of dark matter}
      \author[M. Vogelsberger et al.] {\parbox{18.5cm}{
	  Mark Vogelsberger$^{1}$\thanks{e-mail: mvogelsb@mit.edu}, 
          Jesus Zavala$^{2}$\thanks{Marie Curie Fellow},        
	  Christine Simpson$^{3}$, 
          and Adrian Jenkins$^{4}$
        }\vspace{0.3cm}\\
	$^{1}$ Department of Physics, Kavli Institute for Astrophysics and Space Research, Massachusetts Institute of Technology, Cambridge, MA 02139, USA\\
        $^{2}$ Dark Cosmology Centre, Niels Bohr Institute, University of Copenhagen, Juliane Maries Vej 30, 2100 Copenhagen, Denmark \\
	$^{3}$ Heidelberg Institute for Theoretical Studies, Schloss-Wolfsbrunnenweg 35, 69118 Heidelberg, Germany \\
        $^{4}$ Institute for Computational Cosmology, Durham University, South Road, Durham, UK, DH1 3LE}

\begin{document}
\date{Accepted ???. Received ???; in original form ???}

\pagerange{\pageref{firstpage}--\pageref{lastpage}} \pubyear{2014}

\maketitle

\label{firstpage}

\begin{abstract}
We present the first cosmological simulations of dwarf galaxies, which include
dark matter self-interactions and baryons. We study two dwarf galaxies within
cold dark matter, and four different elastic self-interacting scenarios with
constant and velocity-dependent cross sections, motivated by a new force in the
hidden dark matter sector. Our highest resolution simulation has a baryonic
mass resolution of $1.8\times 10^2\msun$  and a gravitational softening length
of $34\pc$ at $z=0$.  In this first study we focus on the regime of mostly
isolated dwarf galaxies with halo masses $\sim10^{10}\msun$ where dark matter
dynamically dominates even at sub-kpc scales. We find that while the global
properties of galaxies of this scale are minimally affected by allowed
self-interactions, their internal structures change significantly if the cross
section is large enough within the inner sub-kpc region. In these
dark-matter-dominated systems, self-scattering ties the shape of the stellar
distribution to that of the dark matter distribution.  In particular, we find
that the stellar core radius is closely related to the dark matter core radius
generated by self-interactions.  Dark matter collisions lead to dwarf galaxies
with larger stellar cores and smaller stellar central densities compared to the
cold dark matter case. The central metallicity within $1\kpc$ is also larger by
up to $\sim 15\%$ in the former case. We conclude that the mass
distribution, and characteristics of the central stars in dwarf galaxies can 
potentially be used to probe the 
self-interacting nature of dark matter.
\end{abstract}

\begin{keywords}
cosmology: dark matter -- galaxies: halos -- methods: numerical 
\end{keywords}

\section{Introduction}

Low-mass galaxies are arguably the best places to test dark matter (DM) models
since they are dynamically dominated by the DM haloes they are embedded in well
within their inner regions. The kinematical information that is inferred from
Low Surface Brightness galaxies \citep[e.g.][]{Kuzio2008}, nearby field dwarf
galaxies \citep[e.g.][]{deBlok2008,Oh2008} and Milky Way (MW) dwarf spheroidals
(dSphs) \citep[e.g.][]{Walker2011,Amorisco2013}, seem to favour the presence of
$\mathcal{O}(1\kpc)$ dark matter cores with different degrees of certainty. The
former two cases are more strongly established while the latter is still
controversial \citep[e.g.][]{Breddels2013}, which is unfortunate since the MW
dSphs have the largest dynamical mass-to-light ratios and are thus particularly
relevant to test the DM nature.  Although not necessarily related to the
existence of cores, it has also been pointed out that the population of dark
satellites obtained in CDM $N-$body simulations, are too centrally dense to be
consistent with the kinematics of the MW dSphs \citep{Boylan2011, Boylan2012}.
This problem possibly also extends to isolated galaxies
\citep{Ferrero2012,Kirby2014}.

The increasing evidence of lower than expected central DM densities among
DM-dominated systems is a lasting challenge to the prevalent collisionless Cold
Dark Matter (CDM) paradigm.  On the other hand, the low stellar-to-DM content
of dwarf galaxies represents a challenge for galaxy formation models since
these have to explain the low efficiency of conversion of baryons into stars in dwarf galaxies.
It is possible that these two outstanding issues share a common solution rooted
in our incomplete knowledge of processes that are key to understand how
low-mass galaxies form and evolve: gas cooling, star formation and energetic
feedback from supernovae (SNe). In particular, episodic high-redshift gas
outflows driven by SNe have been proposed as a mechanism to suppress subsequent
star formation and lower, irreversibly, the central DM densities
\citep[e.g.][]{Navarro1996, Pontzen2012}.  Although such mechanism seemingly
produces intermediate mass galaxies (halo mass $\sim5-10\times10^{10}\msun$)
with realistic cores and stellar-to-halo mass ratios \citep[][]{Governato2010,
Governato2012}, it is questionable if it is energetically viable for lower mass
galaxies \citep[][]{Penarrubia2012, Garrison-Kimmel2013}. Even though
environmental effects such as tidal stripping might alleviate this stringent
energetic condition in the case of satellite galaxies \citep[][]{Zolotov2012,
Brooks2013,Arraki2014}, the issue of low central DM densities seems relevant 
even for isolated galaxies \citep{Ferrero2012,Kirby2014}. This seems to
indicate that SNe-driven outflows can only act as a solution to this problem if
they occur very early, when the halo progenitors of present-day dwarfs were less
massive \citep{Teyssier2013,Amorisco2013}.  It remains unclear if such systems
can avoid regenerating a density cusp once they merge with smaller, cuspier,
haloes. It is also far from a consensus that the implementation of strong
``bursty'' star formation recipes in simulations, a key ingredient to reduce
central DM densities, is either realistic or required to actually produce
consistent stellar-to-halo mass ratios \citep[e.g.,][]{Marinacci2013}, and
other observed properties. It is therefore desirable, but challenging, to
identify observables that could unambiguously determine whether bursty star
formation histories with a strong energy injection efficiency (into the DM
particles) are realistic or not.

An exciting alternative solution to the problems of CDM at the scale of dwarfs
is that of Self-Interacting Dark Matter (SIDM). Originally introduced by
\cite{SpergelSteinhardt2000},  it goes beyond the CDM model by introducing
significant self-collisions between DM particles. The currently allowed limit
to the self-scattering cross section is imposed more stringently by
observations of the shapes and mass distribution of elliptical galaxies and
galaxy clusters \citep{Peter2012}, and is set at: ${\sigma/m_\chi<1\,{\rm
cm}^2\,{\rm g}^{-1}}$.  DM particles colliding with roughly this cross section
naturally produce an isothermal core with a $\mathcal{O}(1\kpc)$ size in
low-mass galaxies, close to what is apparently observed. SIDM is well-motivated
by particle physics models that introduce new force carriers in a hidden DM
sector \cite[e.g.][]{Feng2009,Feng2010,Arkani-Hamed2009,Aarssen2012,Tulin2013,Cyr-Racine2013,Cline2014},
which predict velocity-dependent self-scattering cross sections. In the case of
massless bosons for instance, the cross section scales as $v^{-4}$ as in
Rutherford scattering. The renewed interest in SIDM has triggered a new era of
high resolution DM-only SIDM simulations: velocity-dependent in
\cite{Vogelsberger2012} (VZL hereafter), and velocity-independent in
\cite{Rocha2012}, that hint at a solution to the CDM problems in low-mass
galaxies. In particular for the MW dSphs, it has been established that the
resultant dark satellites of a MW-size halo are consistent with the dynamics of
the MW dSphs, have cores of $\mathcal{O}(1\kpc)$ and avoid cluster constraints
only if ${0.6\,{\rm cm}^2\,{\rm g}^{-1}\lesssim\sigma/m_\chi\lesssim1.0\,{\rm
cm}^2\,{\rm g}^{-1}}$, or if the cross section is velocity-dependent
\citep[][]{Zavala2013}. Recently, simulations of SIDM models with new light mediators
have shown that is possible to also suppress the abundance of dwarf galaxies due
to the modified early-Universe power spectrum caused by the interactions of the DM with the dark radiation
\citep{Boehm2014,Buckley2014}.

Given the recent success of SIDM, a natural step is to elevate its status to
that of CDM by studying the synergy between baryonic physics and DM
collisionality in a suitable galaxy formation model. So far, this has been
studied only analytically \citep{Kaplinghat2013}, with a focus in more massive
galaxies where baryons dominate the central potential. Interestingly, in this
case, the DM core size is reduced and the central densities are higher compared
to SIDM simulations without the effect of baryons.  In this paper we
concentrate on the regime of dwarf galaxies by pioneering cosmological
hydrodynamical simulations that include the physics of galaxy formation within
a SIDM cosmology.  We compare them with their counterparts (under the same
initial conditions) in the CDM model with the main objective of understanding
the impact of SIDM on the formation and evolution of dwarf galaxies.

\begin{table}
\begin{center}
\begin{tabular}{cccc}
\hline
Name         & $\sigma_T^{\rm max}/m_\chi [{\rm cm}^2\,{\rm g}^{-1}]$     &   $v_{\rm max} [{\rm km}\,{\rm s}^{-1}]$          & allowed?           \\
\hline
\hline
CDM          & --                              & --   & yes                          \\  \hline
SIDM1        & $1$                             & --   & maybe                              \\  \hline
SIDM10       & $10$                            & --   & no                              \\  \hline
vdSIDMa      & $3.5$                           & $30$ & yes                             \\  \hline
vdSIDMb      & $35$                            & $10$ & yes                             \\  \hline
\hline
\end{tabular}
\end{center}
\caption{DM models considered in this paper. CDM is the standard collisionless
model without any self-interaction. SIDM10 is a reference model with a constant
cross section an order of magnitude larger than allowed by current
observational constraints. We note that such a model could still be realized in nature if this
large cross section would only hold over a limited relative velocity range.
SIDM1 is also a model with constant cross section,
which is potentially in the allowed range. vdSIDMa and vdSIDMb have a
velocity-dependent cross section motivated by the particle physics model
presented in \protect\cite{Feng2009, Loeb2011}. These two models are allowed by all
astrophysical constraints, and solve the ``too big to fail'' problem
\protect\citep[see][]{Boylan2011} as demonstrated in VZL.} 
\label{table:ref_points}
\end{table}

This paper is organised as follows. We describe our simulations and the DM models that
we explore in Section~\ref{sims_sec}. We continue with
Section~\ref{visual_sec}, where we give a first visual impression of our dwarf
galaxies. After that we provide more quantitative results in
Section~\ref{global_sec}, where we focus on global properties of the dwarf
galaxies and haloes forming in the different DM models. We study the various
spherically averaged profiles of the dwarfs in Section~\ref{interactions_sec}.
In Section~\ref{sec_inner} we focus on the inner parts of the haloes, where we
expect the largest changes due to self-interactions.  We summarise our results
in Section~\ref{concl_sec}.

\begin{table}
\begin{tabular}{lllllllll}
\hline
name                   & $m_{\rm dm}$    & $m_{\rm baryon}$    & $\epsilon$     & $N_{\rm DM}^{\rm hires}$\\
                       & [$10^2\msun$]   & [$10^2\msun$]       & [${\rm pc}$]   &           \\
\hline
\hline
dA-CDM-B-hi            &  9.7            & 1.8                 & 34.2           & 122,729,602\\
\hline
dA-CDM-B               & 77.5            & 14.8                & 68.5           & 15,353,772\\
dA-SIDM1-B             & 77.5            & 14.8                & 68.5           & 15,353,772\\
dA-SIDM10-B            & 77.5            & 14.8                & 68.5           & 15,353,772\\
dA-vdSIDMa-B           & 77.5            & 14.8                & 68.5           & 15,353,772\\
dA-vdSIDMb-B           & 77.5            & 14.8                & 68.5           & 15,353,772\\
\hline
dA-CDM                 & 77.5            & --                  & 68.5           & 15,353,772\\
dA-SIDM1               & 77.5            & --                  & 68.5           & 15,353,772\\
dA-SIDM10              & 77.5            & --                  & 68.5           & 15,353,772\\
dA-vdSIDMa             & 77.5            & --                  & 68.5           & 15,353,772\\
dA-vdSIDMb             & 77.5            & --                  & 68.5           & 15,353,772\\
\hline
\hline
dB-CDM-B               & 406.2           & 77.4               & 82.2           & 8,196,410\\
dB-SIDM1-B             & 406.2           & 77.4               & 82.2           & 8,196,410\\
dB-SIDM10-B            & 406.2           & 77.4               & 82.2           & 8,196,410\\
dB-vdSIDMa-B           & 406.2           & 77.4               & 82.2           & 8,196,410\\
dB-vdSIDMb-B           & 406.2           & 77.4               & 82.2           & 8,196,410\\
\hline
dB-CDM                 & 406.2           & --                 & 82.2           & 8,196,410\\
dB-SIDM1               & 406.2           & --                 & 82.2           & 8,196,410\\
dB-SIDM10              & 406.2           & --                 & 82.2           & 8,196,410\\
dB-vdSIDMa             & 406.2           & --                 & 82.2           & 8,196,410\\
dB-vdSIDMb             & 406.2           & --                 & 82.2           & 8,196,410\\
\hline
\end{tabular}
\caption{Summary of the simulations. The two dwarf galaxies (dA, dB) are
simulated in CDM and four different SIDM models (SIDM1, SIDM10, vdSIDMa,
vdSIDMb) with (B) and without baryons.  We list the DM particle mass resolution
($m_{\rm dm}$), the cell target mass ($m_{\rm baryon}$), the Plummer-equivalent
maximum physical softening length ($\epsilon$), and the number of DM particles
in the high resolution region. The simulations contain initially the same
number of gas cells in the high resolution region. All models for each halo are
simulated with the same numerical resolution, except for dA-CDM, which was
simulated also with an eight times higher mass resolution (dA-CDM-B-hi) to
check for convergence.}
\label{table:simulations}
\end{table}

\section{Simulations}\label{sims_sec}

To be consistent with our previous work (VZL), we generate zoom-in initial
conditions for two dwarf galaxies from the Millennium-II simulation
\citep[MS-II,][]{Boylan2009}.  The MS-II initial conditions, as the Aquarius
initial conditions studied in our previous SIDM work, use the following
cosmological parameters: $\Omega_{\rm m}=\Omega_{\rm dm} + \Omega_{\rm b} =
0.25$, $\Omega_{\rm b}=0.04$, $\Omega_{\Lambda}=0.75$, $h=0.73$, $\sigma_8=0.9$
and $n_{s}=1$; where $\Omega_{\rm m}$ (with contributions from DM, $\Omega_{\rm
dm}$, and baryons, $\Omega_{\rm b}$) and $\Omega_{\Lambda}$ are the
contributions from matter and cosmological constant to the mass/energy density
of the Universe, respectively, $h$ is the dimensionless Hubble constant
parameter at redshift zero, $n_s$ is the spectral index of the primordial power
spectrum, and $\sigma_8$ is the rms amplitude of linear mass fluctuations in
$8\hmpc$ spheres at redshift zero.

Our simulations include baryons and related astrophysical processes. We employ
the implementation of \cite{Vogelsberger2013} for the moving mesh code {\tt
AREPO} \citep[][]{Springel2010}. We stress that we do not change any parameters
of the model; i.e. we use the same physics parametrisation as the large-scale
simulations in \cite{Vogelsberger2013} and \cite{Torrey2013} and the zoom-in MW
simulations of \citet{Marinacci2013}, \cite{Pakmor2014} and
\cite{Marinacci2014} (with minor modifications). Recently, this model was also
employed to run the Illustris simulation~\citep[][]{Nature2014} and \cite[][for more details]{Vogelsberger2014,
Genel2014}. The model includes: gas cooling and photo-ionisation, star
formation and physics of the interstellar medium, stellar evolution, gas
recycling, chemical enrichment, and kinetic stellar SN feedback.  We note that
we do not include supermassive black holes and AGN feedback in our simulations
since this is not expected to play any role at the mass scale under
consideration.

The implementation for elastic DM self-scattering follows VZL, where DM
interactions are modelled with a Monte Carlo approach. We have ported this
implementation from {\tt GADGET} to {\tt AREPO} without major changes, and we
consider four different elastic SIDM models in this work: SIDM1, SIDM10,
vdSIDMa and vdSIDMb.  The first two have a constant cross section while the
last two have velocity dependent cross sections. The characteristics of the
models are summarised in Table~\ref{table:ref_points}. These models were also
considered in \citet{Vogelsberger2013a} and \citet{Zavala2013} to predict
direct detection signals of SIDM and to constrain the different models using
data from the MW dSphs.

The resolution properties of the simulations are summarised in
Table~\ref{table:simulations}. We simulate two dwarf-scale haloes: dA and dB.
All simulations are carried out with DM-only and with full baryonic physics
(B). The softening length is initially fixed in comoving coordinates, and later
(after $z=7$) limited to a fixed length in physical coordinates, which we list
in Table~\ref{table:simulations} (Plummer-equivalent softening length). We have
performed one higher resolution simulation of dA for the CDM case
(dA-CDM-B-hi). In this first study we will only use this simulation to
demonstrate convergence of our galaxy formation model. The main analysis will
be based on our default resolution. A forthcoming study will exploit
dA-CDM-B-hi and higher resolution SIDM simulations to study the inner profiles
in more detail (Zavala \& Vogelsberger, in prep).

We will show below that the two zoom-in regions are rather different: the dA
environment hosts only one isolated dwarf galaxy, whereas the dB environment
hosts two nearly equal mass isolated dwarf galaxies, which are interacting and
embedded in a rich filament; i.e. the two haloes dA and dB are sampling two
extreme scenarios: a very isolated dwarf with a quiescent formation history,
and an interacting dwarf, which underwent several mergers in the past embedded
in a strong tidal field. Some basic characteristics of the two main isolated
haloes and their galaxies are listed in Table~\ref{table:properties}.  Here we
list virial mass ($M_{\rm 200.crit}$), virial radius ($r_{\rm 200,crit}$),
maximum circular velocity ($V_{\rm max}$), DM mass ($M_{\rm DM}$), stellar mass
($M_\star$), gas mass ($M_{\rm gas}$), V band magnitude ($M_V$), $B-V$ color,
and baryon fraction ($f_b=(M_\star+M_{\rm gas})/M_{\rm 200.crit}$). The two
haloes differ in virial mass by about a factor of $6$, and by a factor of about
$12$ in stellar mass, which also leads to significantly different V band
magnitudes.  The colours ($B-V$) of the two dwarfs are rather similar. We also
include the results for the higher resolution simulation dA-CDM-B-hi to
demonstrate that we achieve excellent convergence in all properties of the
galaxy. We stress that this is a distinct feature of our galaxy formation
model, which was built to lead to convergent results.

Is already clear from Table~\ref{table:properties} that most of the global
quantities of the galactic systems are affected only very little by the DM
nature, and some relative changes are not even systematic with the amplitude of
the scattering cross section due to the stochastic character of star formation
and feedback in our model. The largest systematic differences can be seen
between the CDM case and the extreme SIDM10 model, but even for these two cases the
relative differences are rather small. We will quantify this in more detail
below.

\section{Visual impression}\label{visual_sec}

\begin{figure*}
\centering
\includegraphics[width=0.31\textwidth]{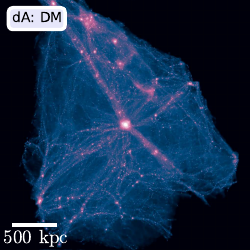}
\includegraphics[width=0.31\textwidth]{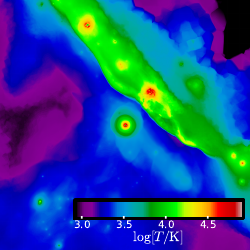}
\includegraphics[width=0.31\textwidth]{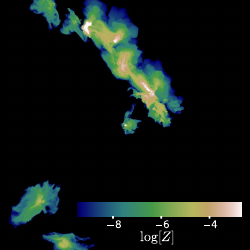}
\includegraphics[width=0.31\textwidth]{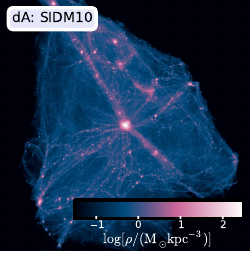}
\includegraphics[width=0.31\textwidth]{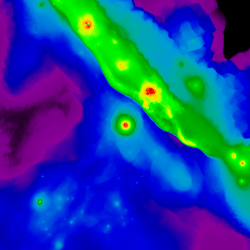}
\includegraphics[width=0.31\textwidth]{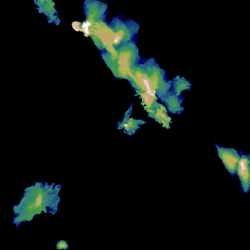}
\includegraphics[width=0.31\textwidth]{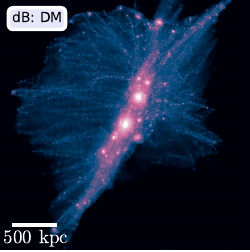}
\includegraphics[width=0.31\textwidth]{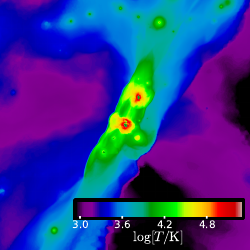}
\includegraphics[width=0.31\textwidth]{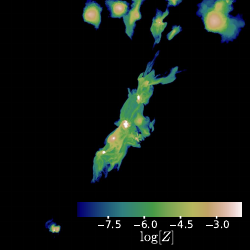}\\
\includegraphics[width=0.31\textwidth]{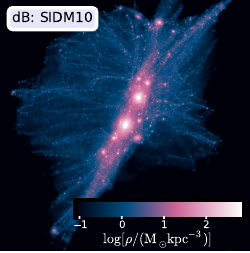}
\includegraphics[width=0.31\textwidth]{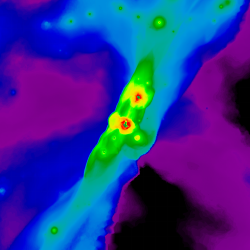}
\includegraphics[width=0.31\textwidth]{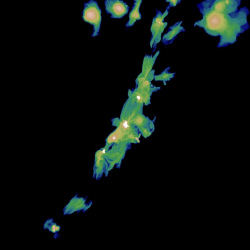}
\caption{Visual overview of the large-scale structure around haloes dA and dB
at $z=0$ for CDM and SIDM with ${\sigma/m_\chi=10\,{\rm cm}^2\,{\rm g}^{-1}}$.
We show from left to right:  DM density, gas temperature, and metallicity
(slice thickness $500\kpc$). The dA dwarf is isolated, whereas dB is embedded
into a rich filament with a few other dwarfs nearby. On large scales, SIDM does
not lead to any significant changes in the DM or gas distribution. The metal
distribution is slightly different indicating that SN-driven outflows operate
slightly differently for CDM than for SIDM due to the modified gravitational
potential in the center. However, this effect is small and stochastic in
nature. The temperature structure shows no visible differences on these
scales.}
\label{fig:fields_z0}
\end{figure*}

\begin{table*}
\begin{tabular}{llllllllllll}
\hline
halo name     & $M_{\rm 200,crit}$   & $r_{\rm 200,crit}$   & $V_{\rm max}$               & $M_{\rm DM}$     & $M_\star$           & $M_{\rm gas}$      &    $M_V$   & $B-V$    & $f_b$  \\
              & [$10^{10}\msun$]     & [${\rm kpc}$]             & [${\rm km}\,{\rm s}^{-1}$]    & [$10^{8}\msun$]  & [$10^{8}\msun$]     & [$10^{8}\msun$]   &            &          &        \\
\hline
dA-CDM-B-hi     &        1.193 	&	 45.841 	&	 49.614 	&	 107.148 	&	 1.478 	&	 12.449 	&	 -15.862 	&	 0.394 	&	 0.117 \\
\hline
dA-CDM-B 	&	 1.198 	&	 45.906 	&	 50.623 	&	 108.773 	&	 1.512 	&	 13.255 	&	 -15.941 	&	 0.382 	&	 0.123 \\
dA-SIDM1-B 	&	 1.193 	&	 45.837 	&	 51.760 	&	 108.631 	&	 1.447 	&	 12.982 	&	 -15.947 	&	 0.371 	&	 0.121 \\
dA-SIDM10-B 	&	 1.164 	&	 45.469 	&	 53.625 	&	 105.578 	&	 1.522 	&	 13.295 	&	 -15.963 	&	 0.386 	&	 0.127 \\
dA-vdSIDMa-B 	&	 1.202 	&	 45.954 	&	 51.982 	&	 109.265 	&	 1.596 	&	 13.147 	&	 -16.006 	&	 0.375 	&	 0.123 \\
dA-vdSIDMb-B 	&	 1.208 	&	 46.030 	&	 50.809 	&	 108.711 	&	 1.502 	&	 13.269 	&	 -15.935 	&	 0.389 	&	 0.122 \\
\hline
\hline
dB-CDM-B 	&	 7.141 	&	 83.223 	&	 83.339 	&	 605.816 	&	 17.712 	&	 118.321 	&	 -18.804 	&	 0.380 	&	 0.190 \\
dB-SIDM1-B 	&	 7.107 	&	 83.097 	&	 86.128 	&	 603.852 	&	 19.142 	&	 115.271 	&	 -18.927 	&	 0.352 	&	 0.189 \\
dB-SIDM10-B 	&	 6.975 	&	 82.577 	&	 87.859 	&	 594.917 	&	 18.131 	&	 114.493 	&	 -18.793 	&	 0.345 	&	 0.190 \\
dB-vdSIDMa-B 	&	 7.136 	&	 83.206 	&	 86.251 	&	 604.041 	&	 17.977 	&	 115.789 	&	 -18.738 	&	 0.372 	&	 0.187 \\
dB-vdSIDMb-B 	&	 7.192 	&	 83.425 	&	 83.092 	&	 608.296 	&	 17.559 	&	 117.623 	&	 -18.731 	&	 0.390 	&	 0.188 \\
\hline
\end{tabular}
\caption{Basic properties of the simulated dwarf galaxies dA and dB. The
different columns list: virial mass ($M_{\rm 200,crit}$), virial radius
($r_{\rm 200,crit}$), maximum circular velocity ($V_{\rm max}$), DM mass
($M_{\rm DM}$), stellar mass ($M_\star$), gas mass ($M_{\rm gas}$), V band
magnitude ($M_V$), $B-V$ color, and baryon fraction ($f_b=(M_\star+M_{\rm
gas})/M_{\rm 200.crit}$). dB is about six times more massive than dA.
Differences in the DM model do not lead to any significant changes in the
global galaxy properties listed here. The dA-CDM-B-hi results demonstrate that
our galaxy formation model leads to excellent convergence of the baryonic
characteristics.}
\label{table:properties}
\end{table*}

\begin{figure*}
\centering
\includegraphics[width=0.24\textwidth]{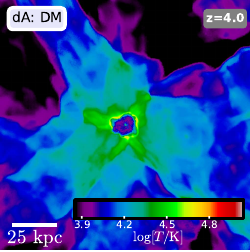}
\includegraphics[width=0.24\textwidth]{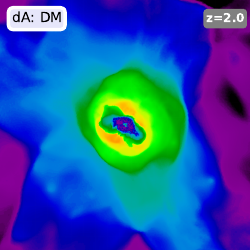}
\includegraphics[width=0.24\textwidth]{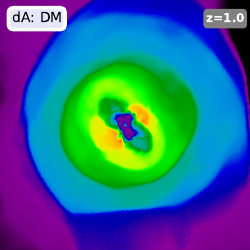}
\includegraphics[width=0.24\textwidth]{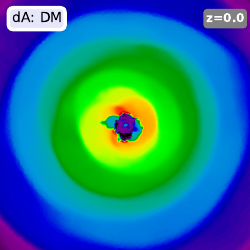}
\includegraphics[width=0.24\textwidth]{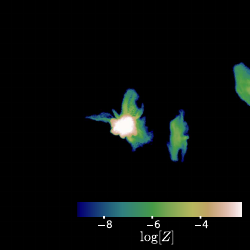}
\includegraphics[width=0.24\textwidth]{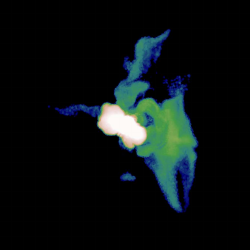}
\includegraphics[width=0.24\textwidth]{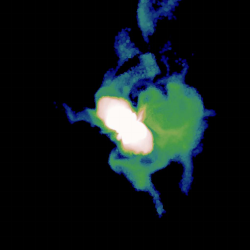}
\includegraphics[width=0.24\textwidth]{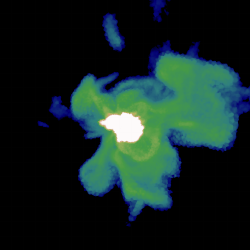}
\includegraphics[width=0.24\textwidth]{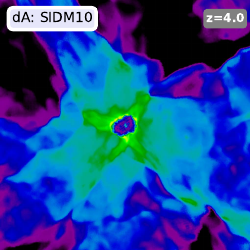}
\includegraphics[width=0.24\textwidth]{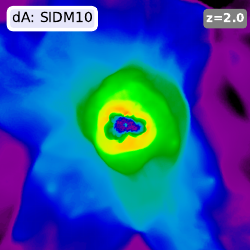}
\includegraphics[width=0.24\textwidth]{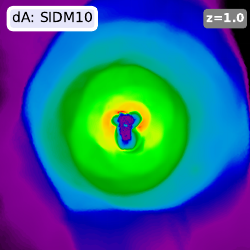}
\includegraphics[width=0.24\textwidth]{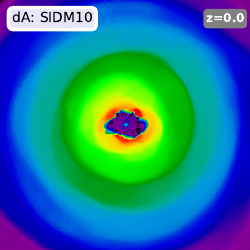}
\includegraphics[width=0.24\textwidth]{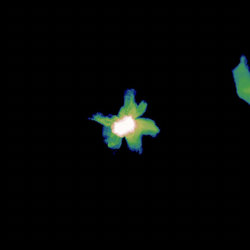}
\includegraphics[width=0.24\textwidth]{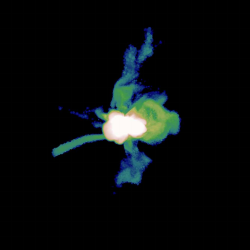}
\includegraphics[width=0.24\textwidth]{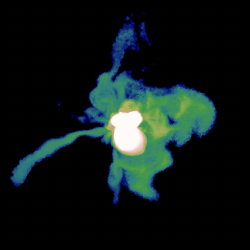}
\includegraphics[width=0.24\textwidth]{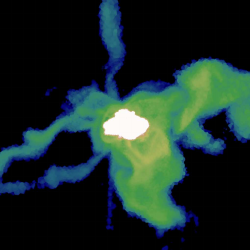}
\caption{Redshift evolution of gas properties of dwarf dA at $z=4,2,1,0.5,0$
(left to right).  We show gas temperature (top), and gas metallicity (bottom)
in slices of thickness $25\kpc$. At $z=4$, the temperature and metallicity
distributions look similar in CDM and SIDM, because SIDM collisions modify the
DM potential only towards later times. This then also induces changes in the
baryonic evolution. The outflows seen in the metallicity maps clearly deviate
between CDM and SIDM. Also the inner temperature structure is affected by
this.}
\label{fig:dA_fields_many_z}
\end{figure*}

We first give some visual impressions of the simulated region at $z=0$ in
Figure~\ref{fig:fields_z0} (left dA, right dB):  from left to right we show the
projected DM density, the gas temperature, and the gas metallicity.  We focus
here only on the most extreme models in terms of self-scattering cross section,
CDM (top) and SIDM10 (bottom). It is clear that the environments of dA and dB
are very different: the dA halo is very isolated, whereas dB lives in a rich
filamentary structure with other haloes surrounding it.  The impact of SIDM on
these scales is minimal: even the extreme SIDM10 model with a cross section
$10$ times larger than observationally allowed, does not alter any of the
fields in a visible way on large scales.  Perhaps the most visible effect in
the DM distribution is a slight decrease in the abundance of (sub)haloes. We
will not quantify this here in detail, but we note that a similar effect was
already found in VZL for the subhalo abundance of MW-like DM haloes. However,
this effect is only visible for very large cross sections, which are
observationally excluded, and is negligible for allowed models.  More
interesting is the fact that the modified gravitational potential of the inner
region of the dwarf through its evolution leads to a different distribution of
SN-driven gas outflows (clearer at smaller scales, see Figure
\ref{fig:dA_fields_many_z}). This effect is visible in the metallicity
projections, where some slight differences are noticeable even on the large
scales shown here.  However, the effect on such large scales is very small, and
it is therefore unlikely that the distribution of baryons on these scales can
be used to probe the DM nature.

The build-up of the dA dwarf can be inspected in
Figure~\ref{fig:dA_fields_many_z}, where we show the evolution at
five redshifts ($z=4, 2, 1, 0.5, 0$).  Here we focus only on the
evolution of the gas properties: gas temperature, and gas metallicity.
Furthermore, we show a much smaller region around the halo compared to
Figure~\ref{fig:fields_z0} (as indicated by the scale). It is clear that halo
dA has essentially grown in isolation since $z=4$, while halo dB has had a
violent merger history (not explicitly shown here).  Notice that by $z=4$ there are only very minor
differences between CDM (top panels) and SIDM10 (bottom panels), most visible in the metallicity
distribution. This is because DM collisions are only relevant at lower
redshifts once the densities in the centers of haloes are high enough for
scatterings to occur. 

The further evolution demonstrates that small variations in the inner halo DM
potential due to DM collisions can alter the subsequent
evolution of the galaxy. This is more spectacularly seen in the divergent
history of outflows driven by SNe, which are clearly visible in the metallicity maps. However, this should be interpreted with
care due to the stochastic nature of star formation and wind generation in our
implementation. It seems that this is the main driver, e. g., by looking at the
SFR there is no clear correlation with the amplitude of the cross section (see Figure \ref{fig:globalprop}). The conclusion seems
to be that globally, this stochastic nature makes it impossible to distinguish
the different DM models, which is why one needs to focus on the inner regions of
the dwarfs to look for clues of DM collisions.
We demonstrate below that baryonic characteristics of the inner galaxies (within $\sim 1\kpc$) are closely
related to the DM model. This is, of course, not surprising since the largest
effect of SIDM occurs in the center of galaxies through core formation.

\section{Global properties: comparison to observations}\label{global_sec}

We now describe the global properties, integrated over the whole galaxy, of our
simulated dwarfs and compare some of them with observations of dwarf galaxies.
Our intention in this work is not a detailed observational comparison, but
rather to study the impact of SIDM on the baryonic component.  Nevertheless, we
would like to quantify how ``realistic'' our dwarf galaxies are in terms of
their baryonic content. The comparison here is rather limited since our dwarf
sample is very small, and because the model we are using is actually ``tuned''
for a slightly different cosmology, and for a somewhat larger mass scale
\citep[see][for details]{Vogelsberger2013}. With these caveats in mind, we
compare the two dwarfs to a few observations below.  This section will also
demonstrate that the impact of SIDM on global and integrated galaxy properties
is typically negligible at the mass scales we explore here.  

\begin{figure*}
\centering
\includegraphics[width=0.33\textwidth]{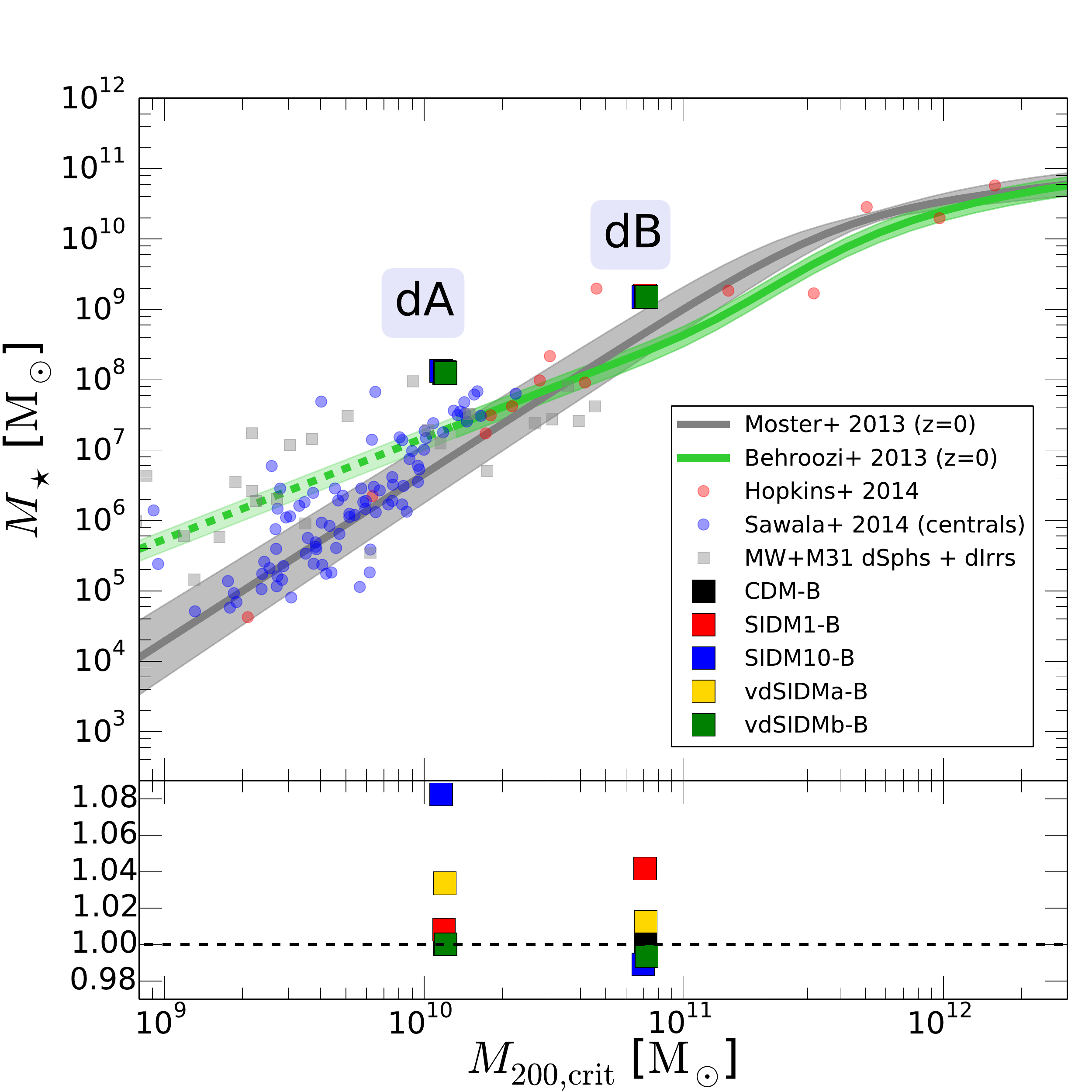}
\includegraphics[width=0.33\textwidth]{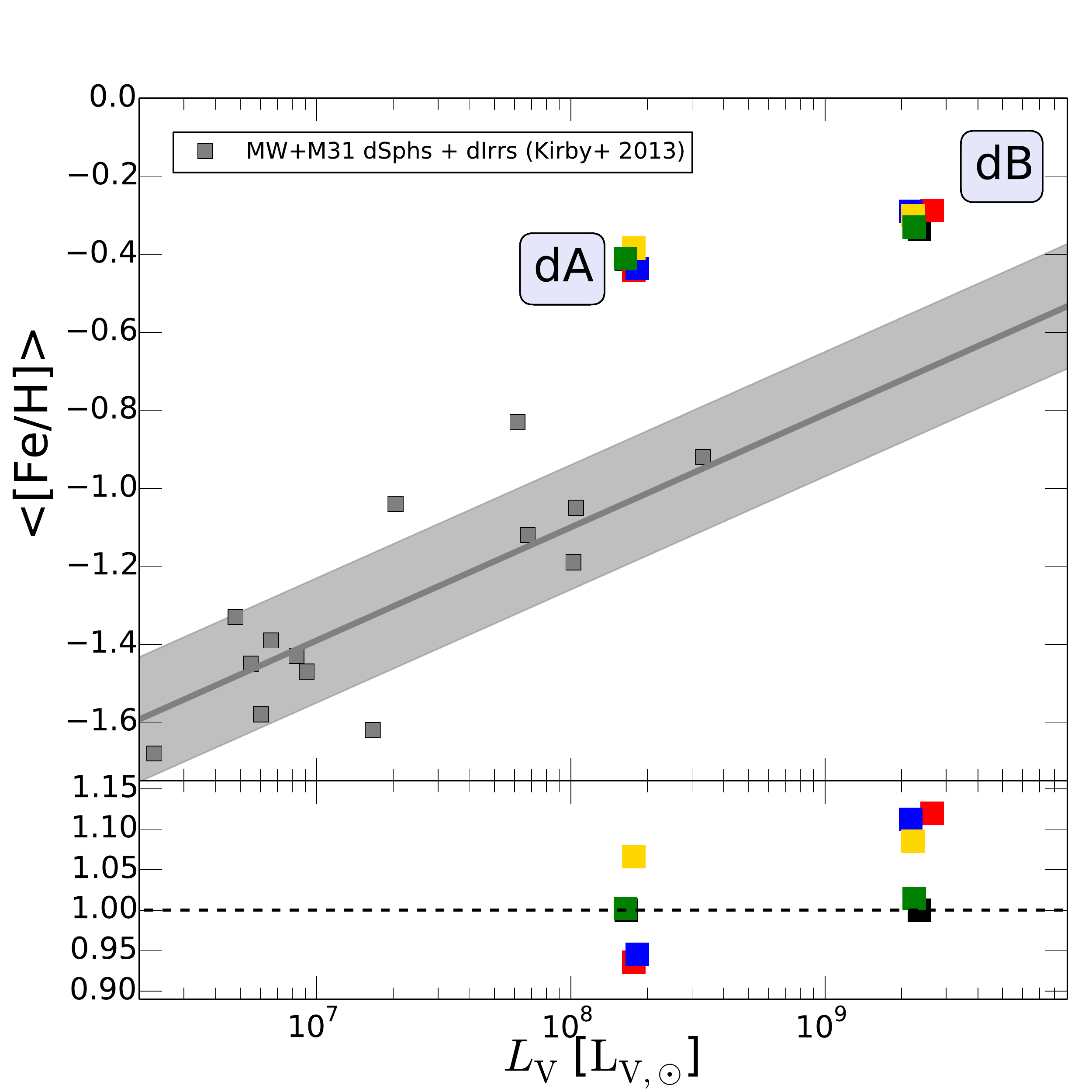}
\includegraphics[width=0.33\textwidth]{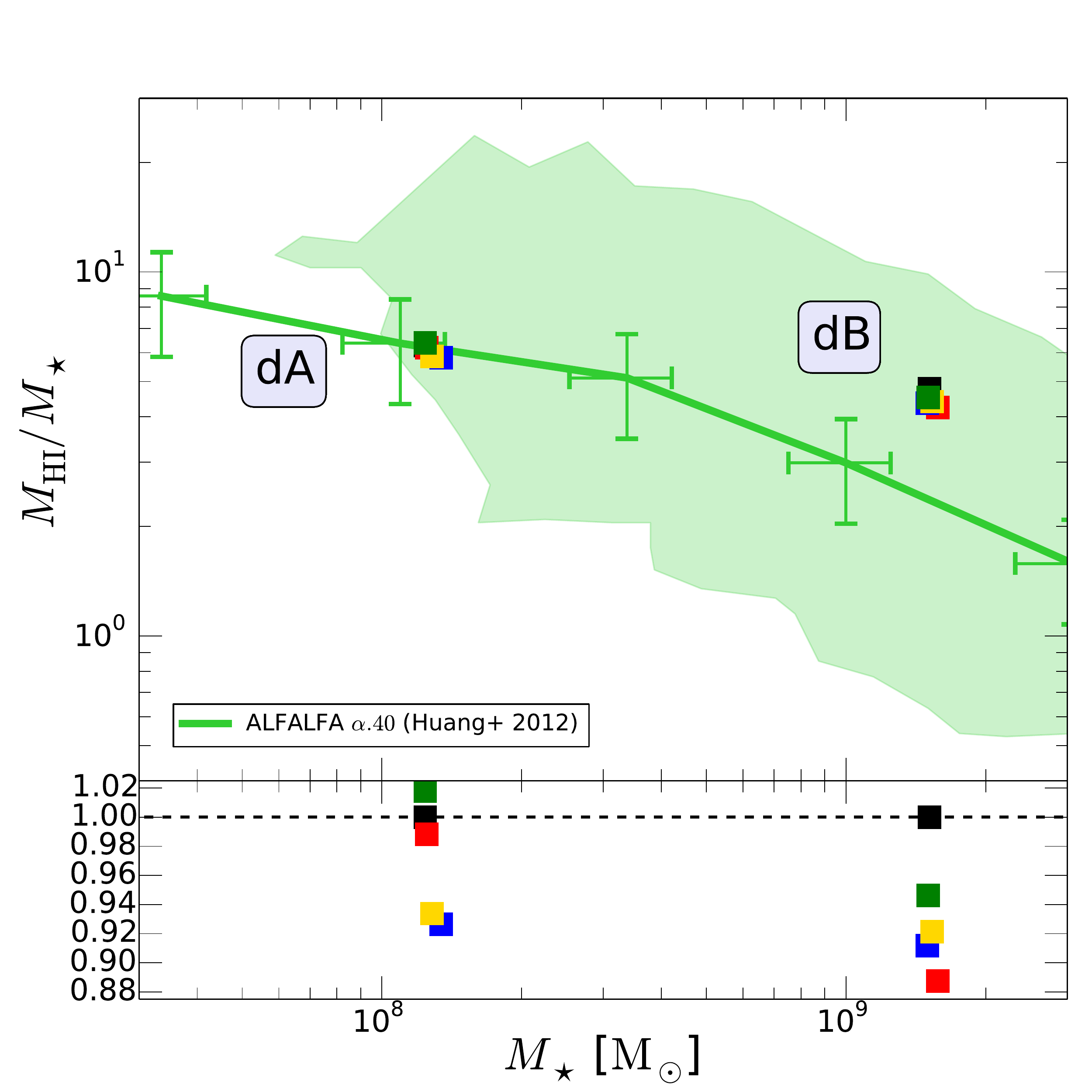}
\includegraphics[width=0.33\textwidth]{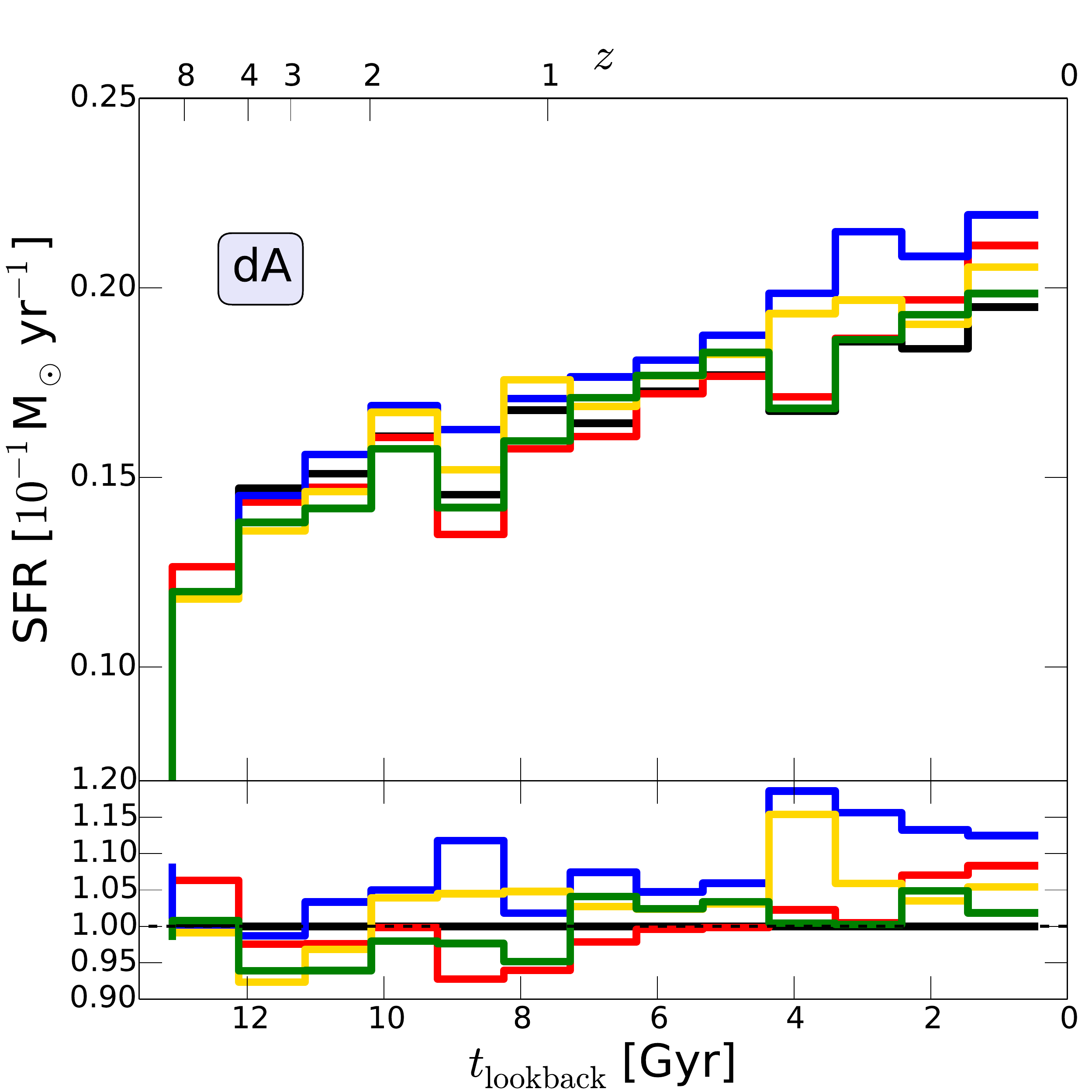}
\includegraphics[width=0.33\textwidth]{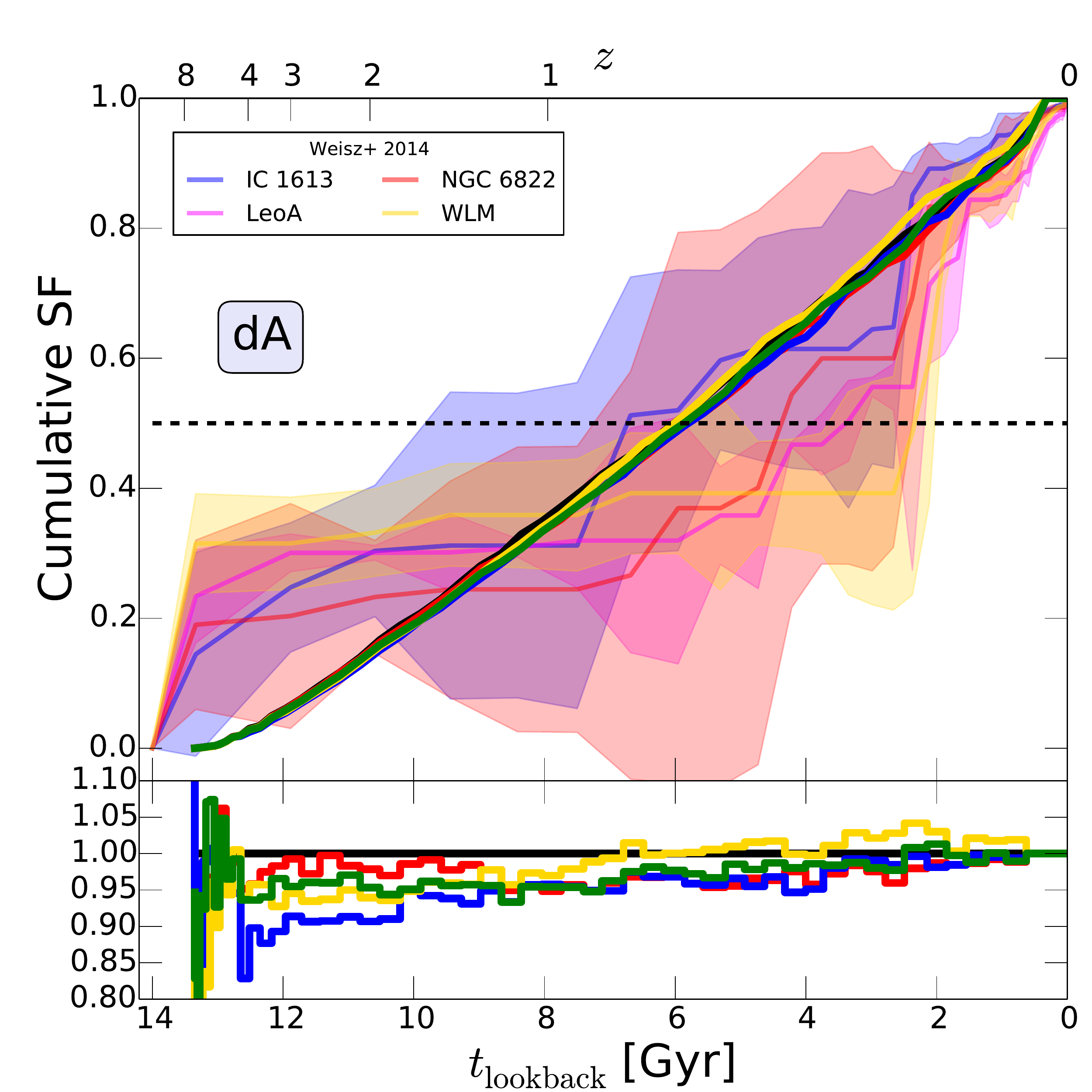}
\includegraphics[width=0.33\textwidth]{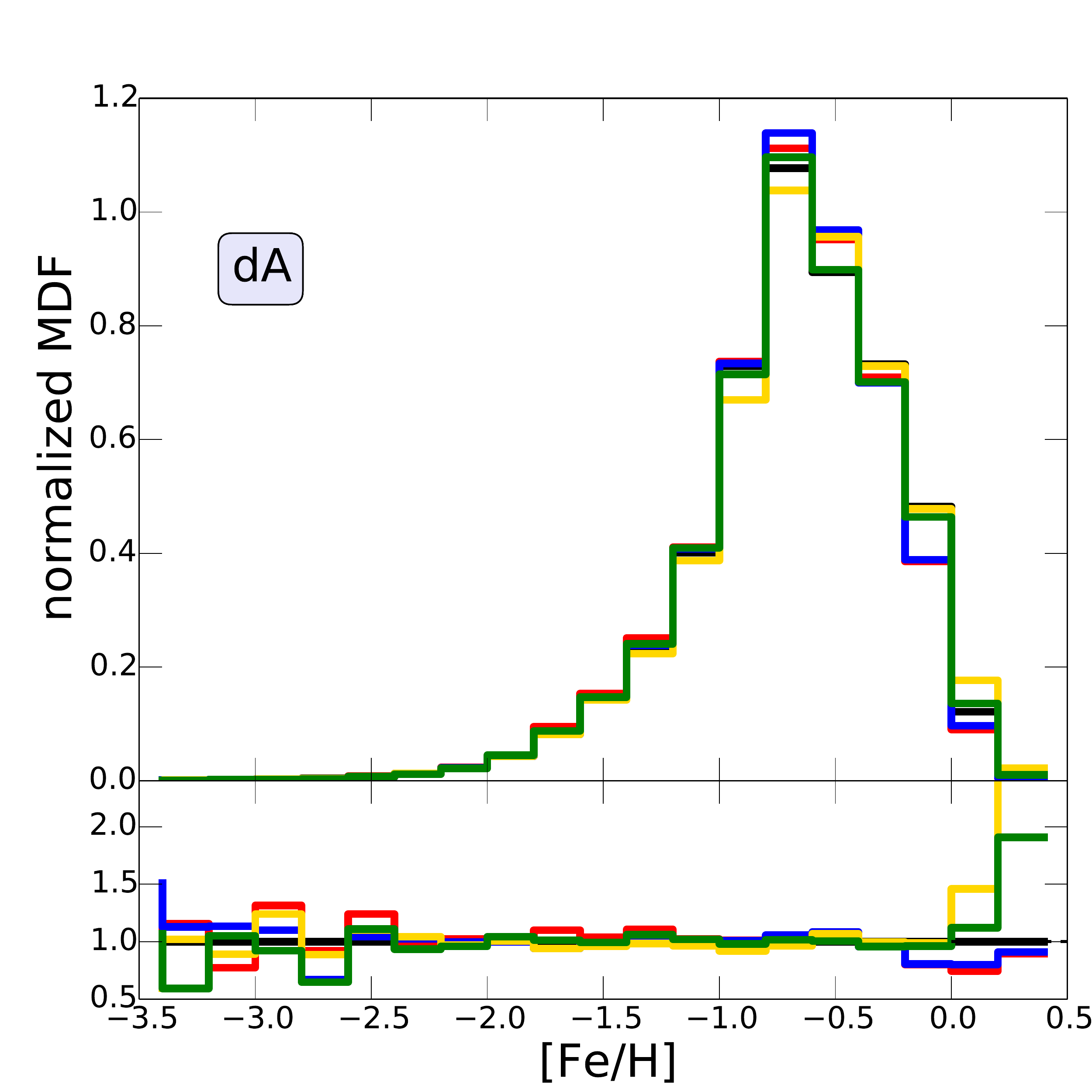}
\includegraphics[width=0.33\textwidth]{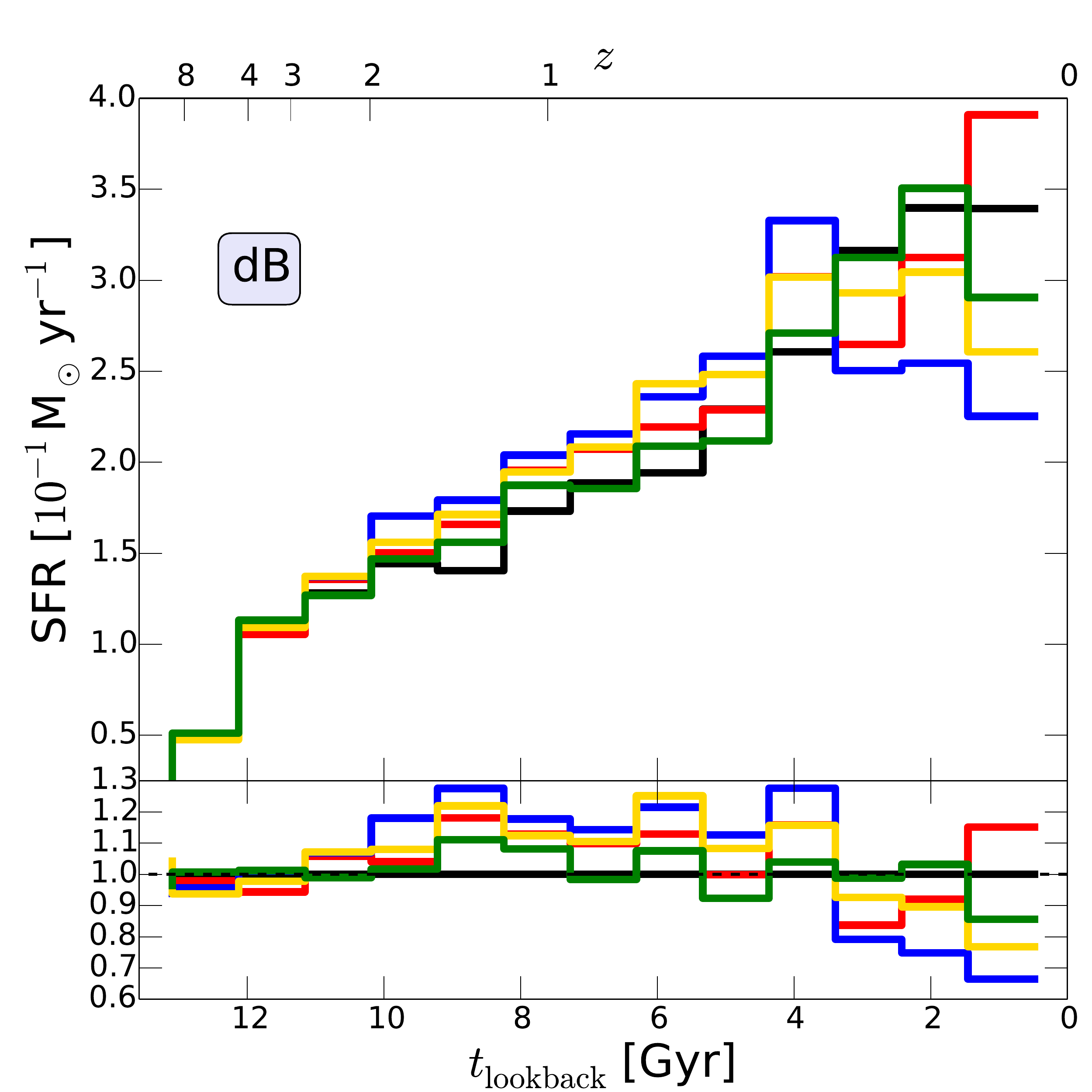}
\includegraphics[width=0.33\textwidth]{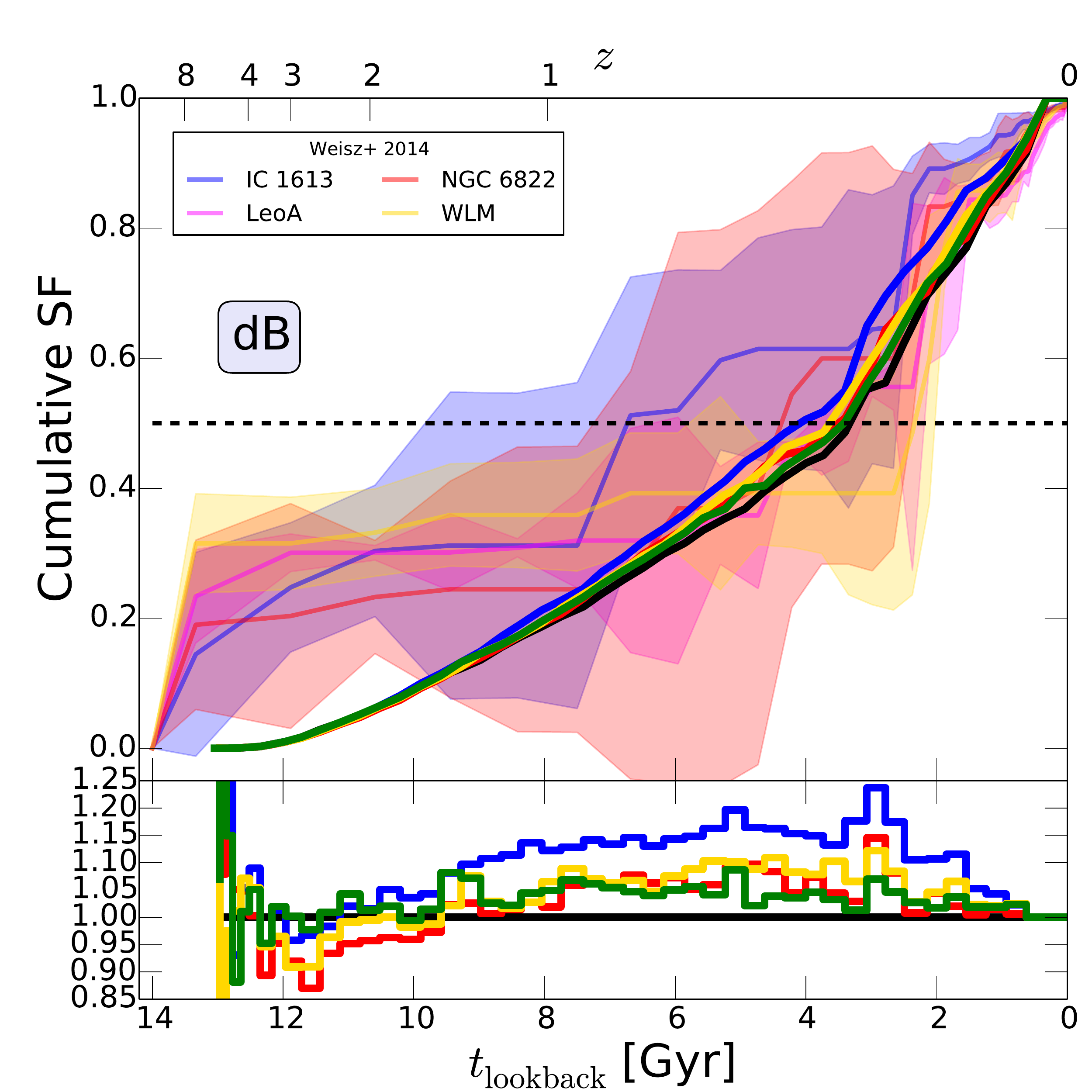}
\includegraphics[width=0.33\textwidth]{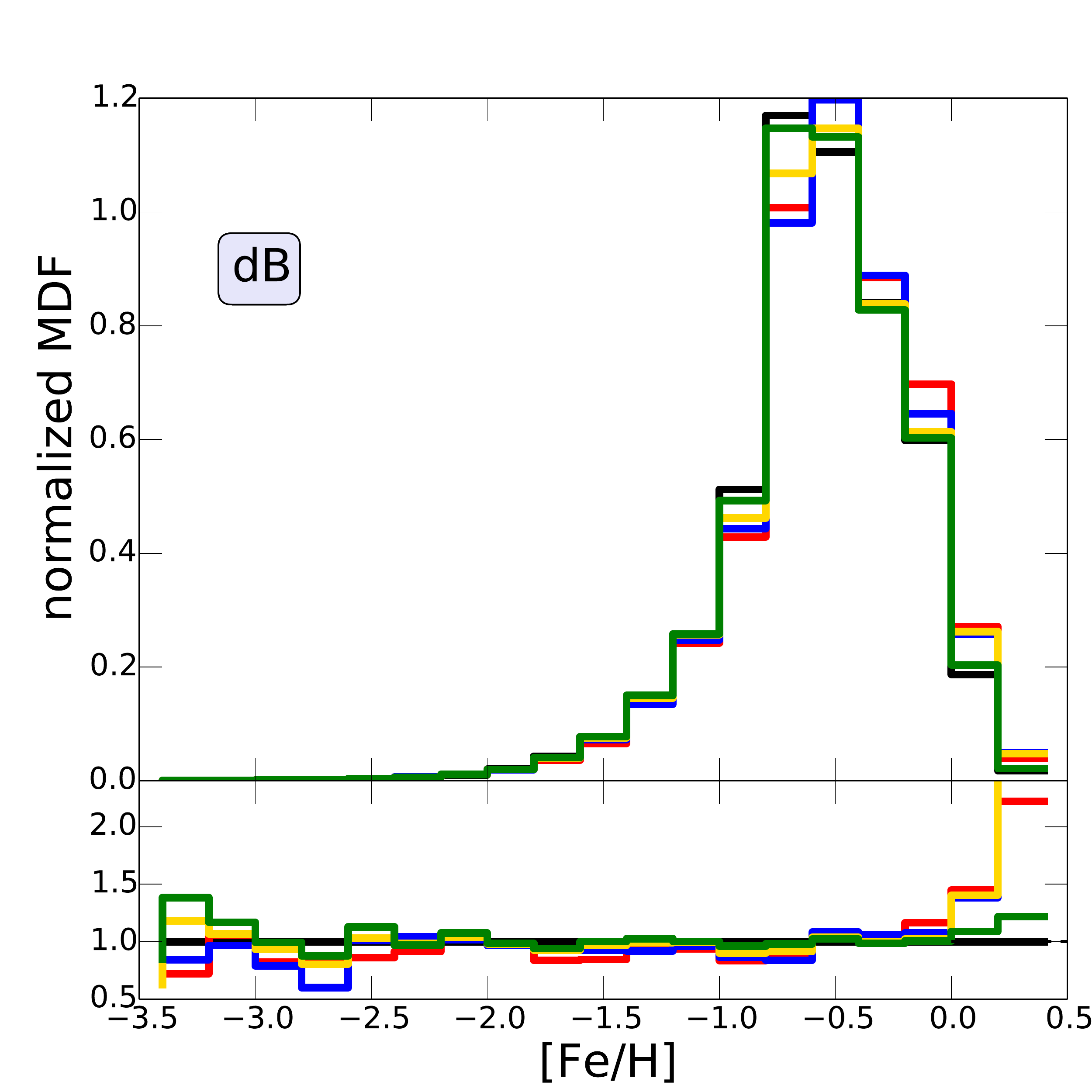}
\caption{A selection of global properties of the simulated galaxies at $z=0$.
Top three panels (left to right): the stellar mass as a function of halo mass
compared to recent abundance matching results from~\protect\cite{Moster2013}
and~\protect\cite{Behroozi2013}, to simulation results from \citet{Hopkins2014} and \citet{Sawala2014}, and to 
observations of local group dwarfs; the metallicity luminosity relation compared
to observations from~\protect\cite{Kirby2013}; and the {\HI} mass richness
relation compared to observations from~\protect\cite{Huang2012}.  The other six
panels show for the two haloes (top: dA, bottom: dB): the star formation
history in bins of $0.5\Gyr$ (left); the cumulative star formation history
(middle) compared to four local group dwarfs taken from \protect\cite{Weisz2014}, which have similar stellar masses and are reasonably isolated based
on the host distance, and the stellar metallicity distribution function
(right).  In the bottom of each panel we show the ratio of each model to that
of CDM. The nature of DM (CDM vs.  SIDM) does not lead to any significant and
systematic differences in the global properties of galaxies. Typical changes
are of the order of $10\%$ at most.}
\label{fig:globalprop}
\end{figure*}

We first study the star formation efficiency by measuring the total stellar
mass within $r<r_\star$ at $z=0$, where $r_\star$ is twice the stellar
half-mass radius, which we define as our fiducial galaxy radius~\citep[see][for
details]{Vogelsberger2013}.  The result is shown in the upper left panel of
Figure~\ref{fig:globalprop} for the different scenarios according to the legend
for the less massive halo (dA) and the more massive halo (dB).  

We compare our results to the empirical $M_{\star}-M_h$ relation obtained using the
abundance matching technique for observed galaxies at $z=0$
\citep{Behroozi2013, Moster2013}.  Compared to these, halo dA has formed too
many stars while halo dB has the right stellar mass content at $z=0$ being
within the observational $1\sigma$ uncertainties. We also include recent
simulation results from~\cite{Hopkins2014} and \cite{Sawala2014}
(simulated in a local group environment), along with
observational data of local group dwarfs (MW+M31 dSphs + dIrrs)~\citep[taken
from][]{Cote2000, McGaugh2005, Woo2008, Penarrubia2008, Stark2009, Oh2011,
Misgeld2011, Ferrero2012, Tollerud2012}. Considering all this simulation and 
observational data, it seems that the stellar mass of dA, albeit in the high end, is
not unreasonable. Nevertheless, we still need a larger simulation sample
of galaxies in that mass range to test how reasonable is the galaxy formation model we have used.  Since this is
not the main focus of the current paper, we leave it for future studies,
concentrating instead on the contrast between the different DM models in the following sections.

In the left panels in the second and third rows of Figure~\ref{fig:globalprop},
we show the star formation rates as a function of look-back time in $0.5$~Gyr
bins. For this, we consider all stellar particles which belong to the galaxy
($r<r_\star$) at $z=0$.  With this age resolution, our model does not lead to a
very bursty star formation history, although the time evolution is also not
completely smooth.  We stress, that there is currently no undisputed direct
observational evidence for bursty star formation histories for dwarfs like the
ones simulated here. It remains to be seen which distribution of star formation
histories is actually realised in Nature.  Nevertheless, we should note that a
recent analysis by \citet{Kauffmann2014} seems to give convincing evidence that
most $M_\star\sim10^8\msun$ galaxies suffer ongoing bursts of star formation
with a typical duration ($\Delta t_{\rm burst}$) of the order of the
characteristic dynamical time of the galaxy ($\Delta t_{\rm dyn}$).  Although
this might suggest that the gas outflows from these bursts could change the DM
distribution, it is not clear how efficient this would be since the highly
efficient regime occurs only once $\Delta t_{\rm burst}\ll\Delta t_{\rm dyn}$.  

The star formation rates of the two dwarfs show a slightly different behaviour:
the rate of dA is fluctuating around a moderately non-evolving mean, whereas dB
has a more significantly increasing mean. Most importantly, none of our dwarfs
have an exponentially declining star formation history. These trends are
actually similar to models with more explicit stellar feedback \citep[see][for
example]{Hopkins2014}.  The middle panels in the second and third rows in
Figure~\ref{fig:globalprop} show the fractional cumulative star formation,
i.e., the fraction of stellar mass formed before the indicated time.  We
compare the simulation results to a few dwarf galaxies based on a sample from
\cite{Weisz2014} of four local group dwarf irregulars lying in a
similar stellar mass range as the simulated ones and that are not disturbed too much
by the tidal field of the MW and Andromeda (see e.g.  Figure~1 of
\cite{Leaman2012} for a visual impression). The cumulative star formation
histories of our dwarf galaxies do not deviate strongly from the observational
data. Notice how the observed dwarfs seem to have larger star formation rates
at very early ($z>4$) and very late times ($t_{\rm lookback}\lesssim4\Gyr$)
compared to our isolated dwarf (dA). The seemingly good agreement with our
systems dB in the late time regime might indicate that interactions are
responsible for the late time surge of star formation.  However, this is only
speculative since the observed dwarfs are in relative isolation today.  The
high redshift regime might be related to a period of star formation before
reionisation and thus, to a scenario where the observed dwarfs come from
progenitors that collapsed earlier than the haloes we simulate here.  

In the bottom of each panel of  Figure~\ref{fig:globalprop}, we show the ratios
of the different DM models with respect to the CDM prediction.  The total
stellar masses have variations of the order of only $\sim10\%$ and, as the star
formation histories suggest, these are likely related to the stochastic nature
of star formation (and SNe-driven winds) in our simulations. Looking at the star
formation rates for instance, it is clear that there is no trend with the
amplitude of the scattering cross section (noticeable with more clarity by
comparing the red, SIDM1, and blue, SIDM10, lines in both panels).
Nevertheless, the total amount of stars tends to either be very identical to
the stellar mass formed in CDM, or a few percent higher according to the upper
left panel. 

The upper right panel of Figure~\ref{fig:globalprop} shows the neutral hydrogen
{\HI} richness relation for our two galaxies compared to data from the ALFALFA
survey ($40\%$ of the catalogue of \citealt{Huang2012}).  Both simulated galaxies
lie within the the observed distribution, although halo dB is more {\HI}-rich
than the observed mean.  The second panel in the top row shows the $V$-band
luminosity metallicity relation, where a comparison with the compilation of
data for dSphs and dIrrs presented in \cite{Kirby2013} is also shown. These two
types of galaxies seem to obey the same relation. Our simulated galaxies are
slightly too metal-rich compared to observations, particularly for the smallest
dwarf. Finally, the right panels in the second and third rows of
Figure~\ref{fig:globalprop} show the metallicity distribution functions. 

\begin{figure*}
\centering
\includegraphics[width=0.33\textwidth]{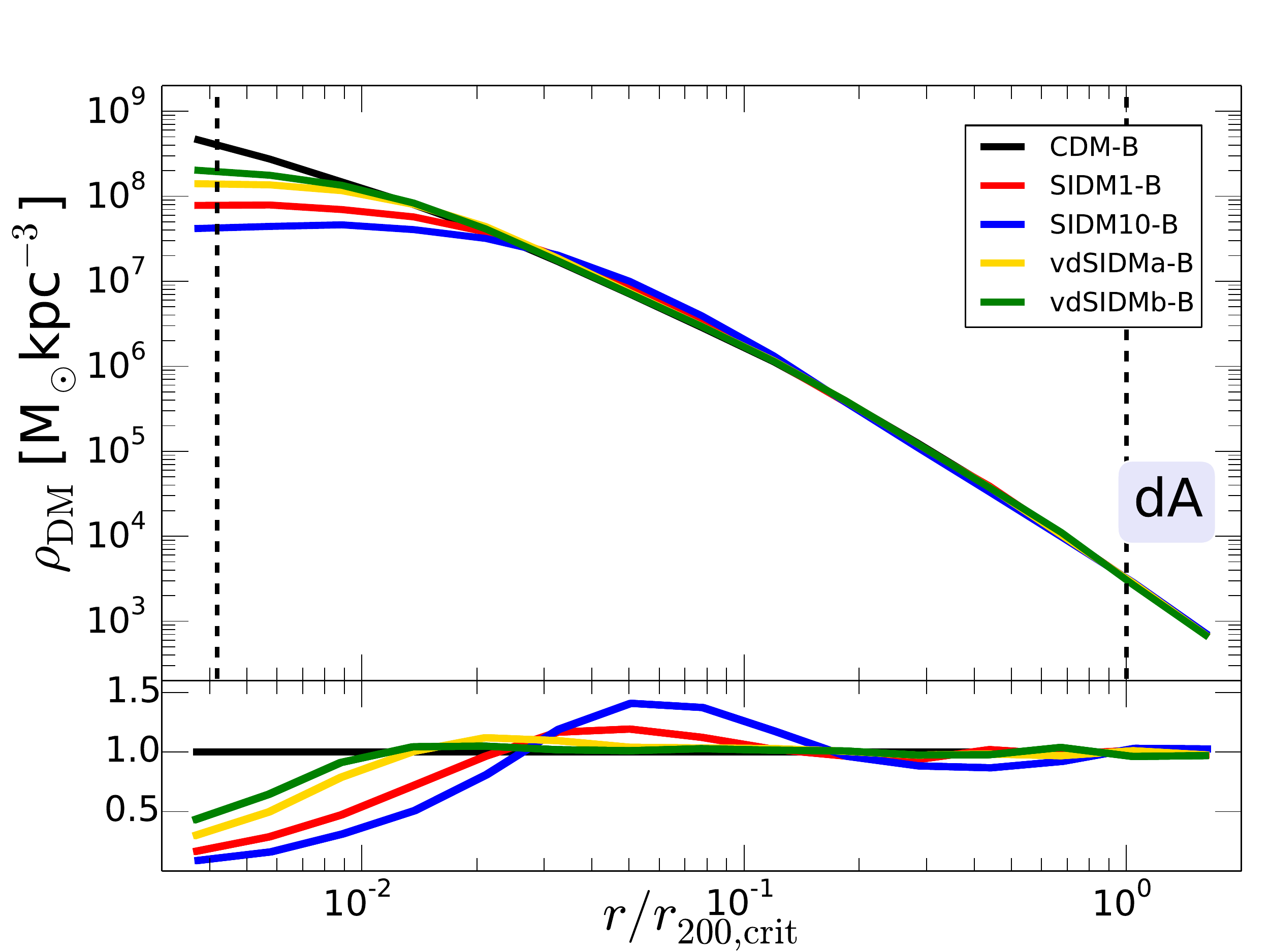}
\includegraphics[width=0.33\textwidth]{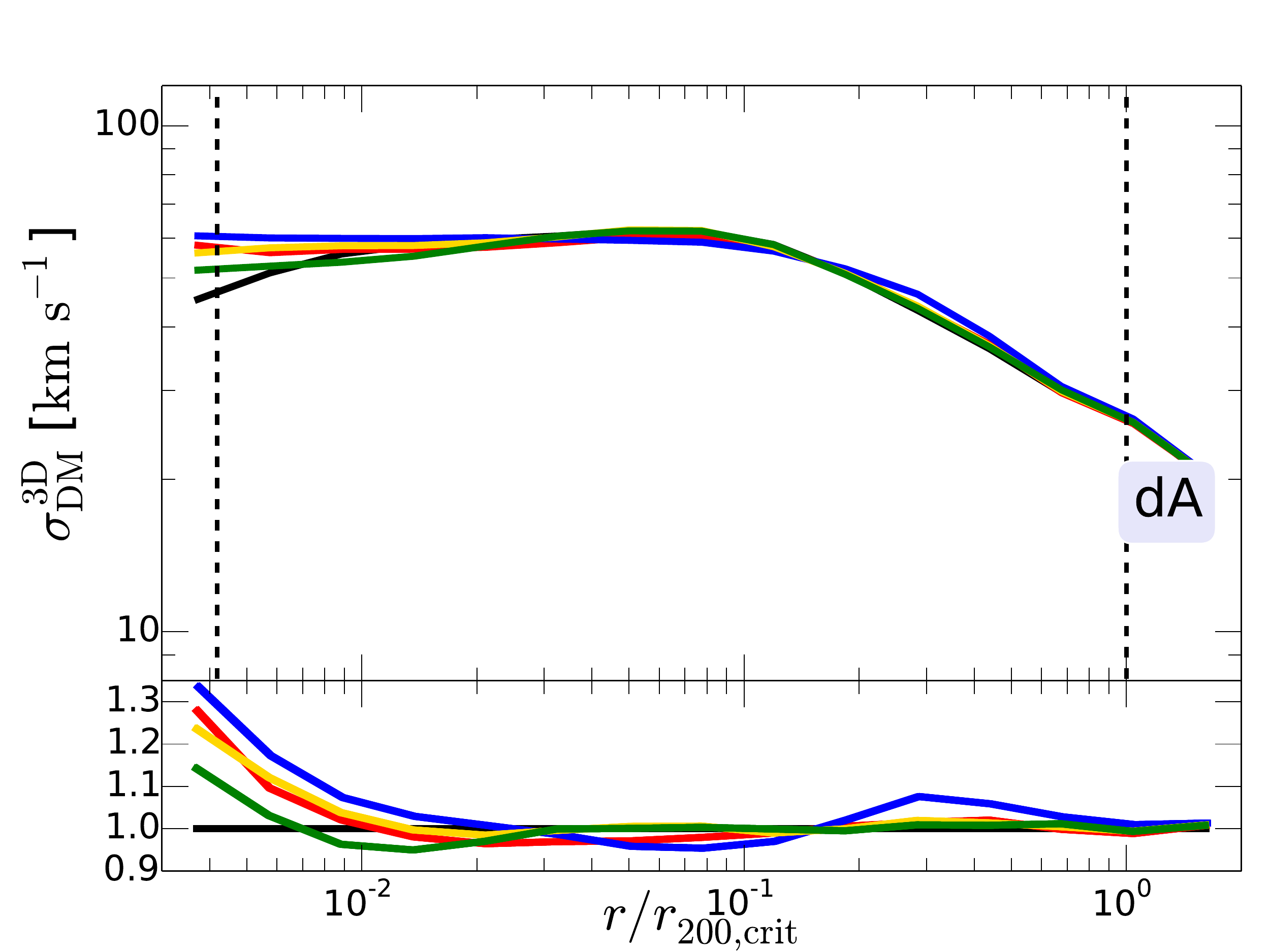}
\includegraphics[width=0.33\textwidth]{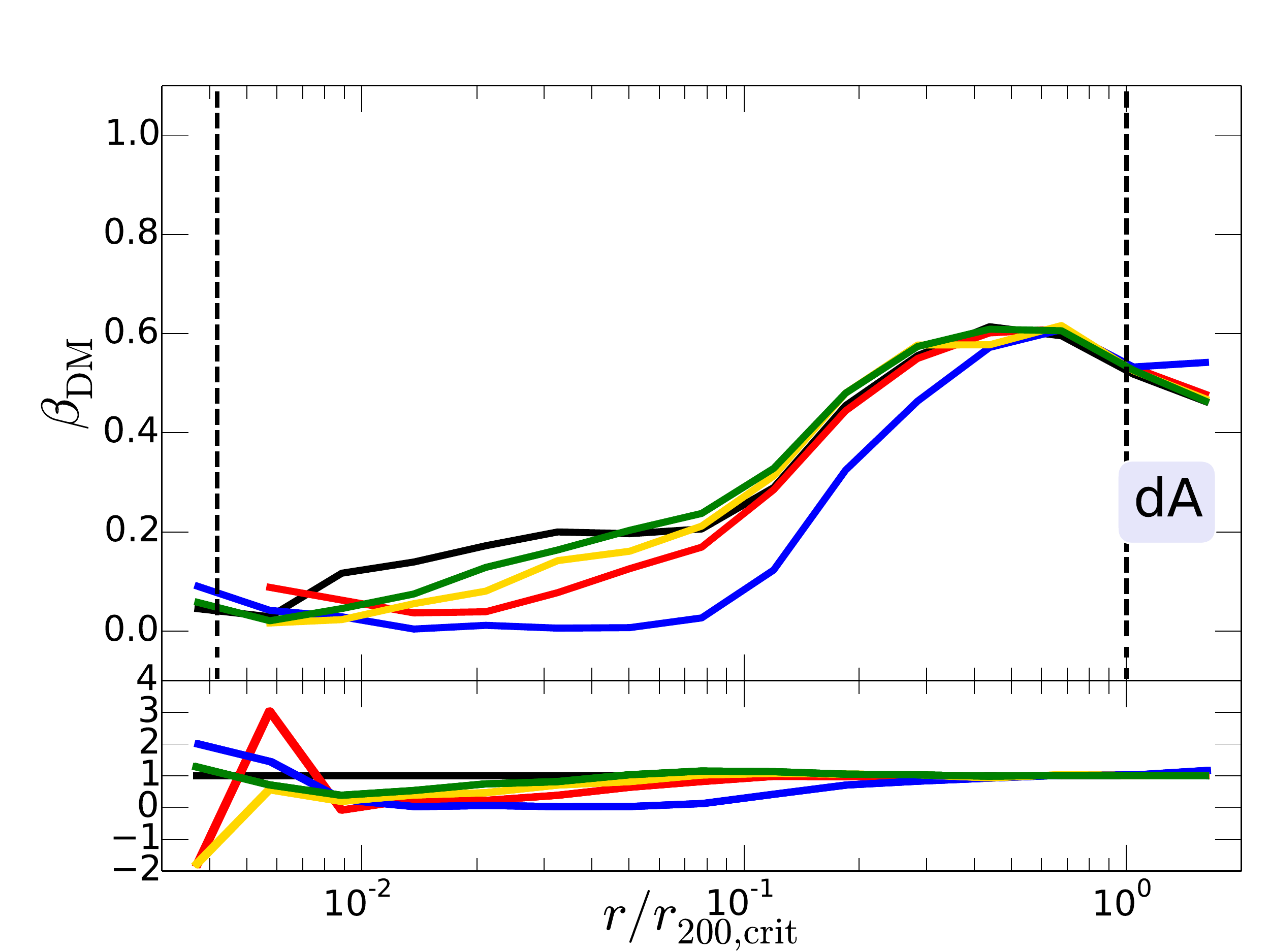}
\includegraphics[width=0.33\textwidth]{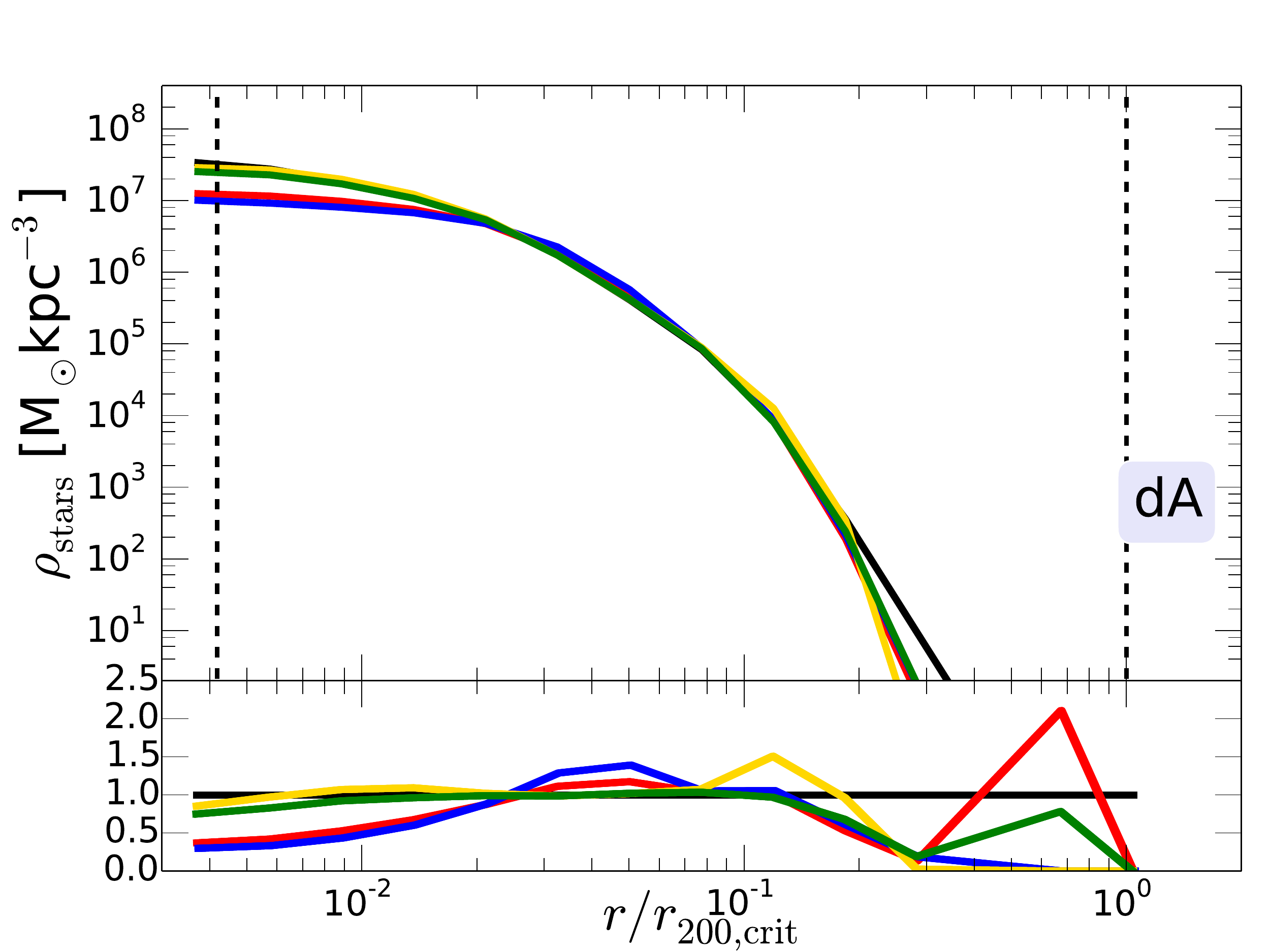}
\includegraphics[width=0.33\textwidth]{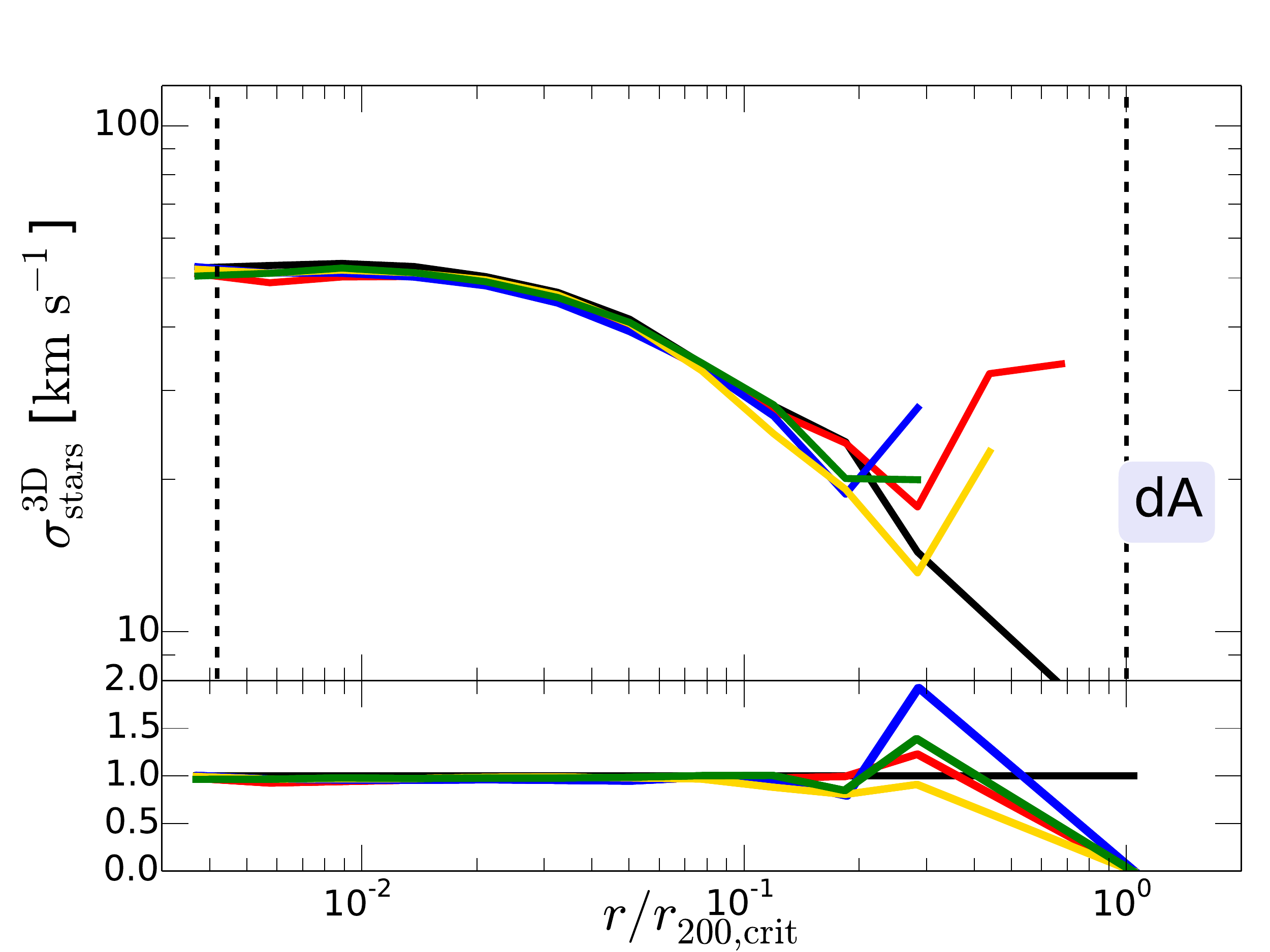}
\includegraphics[width=0.33\textwidth]{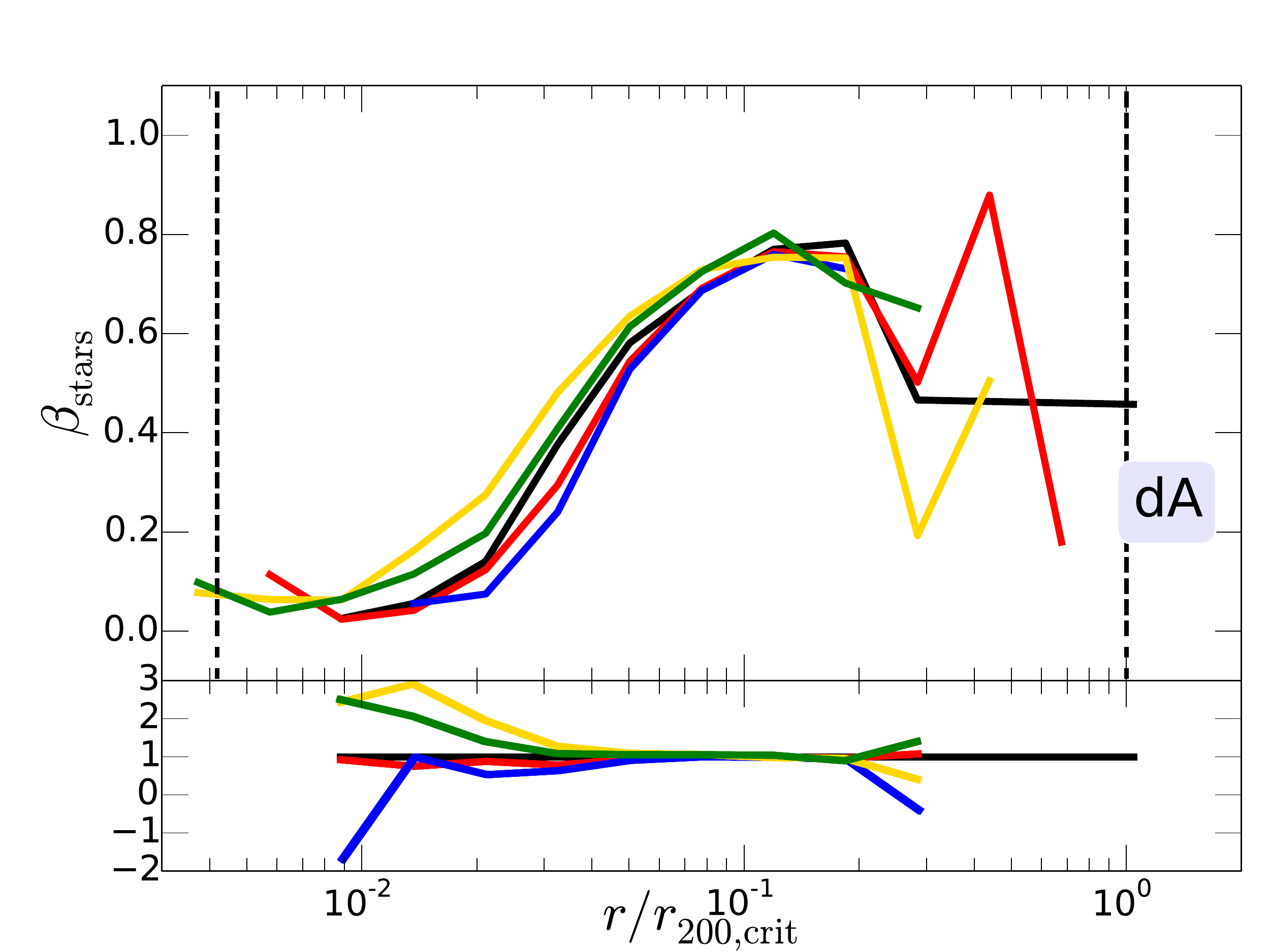}
\includegraphics[width=0.33\textwidth]{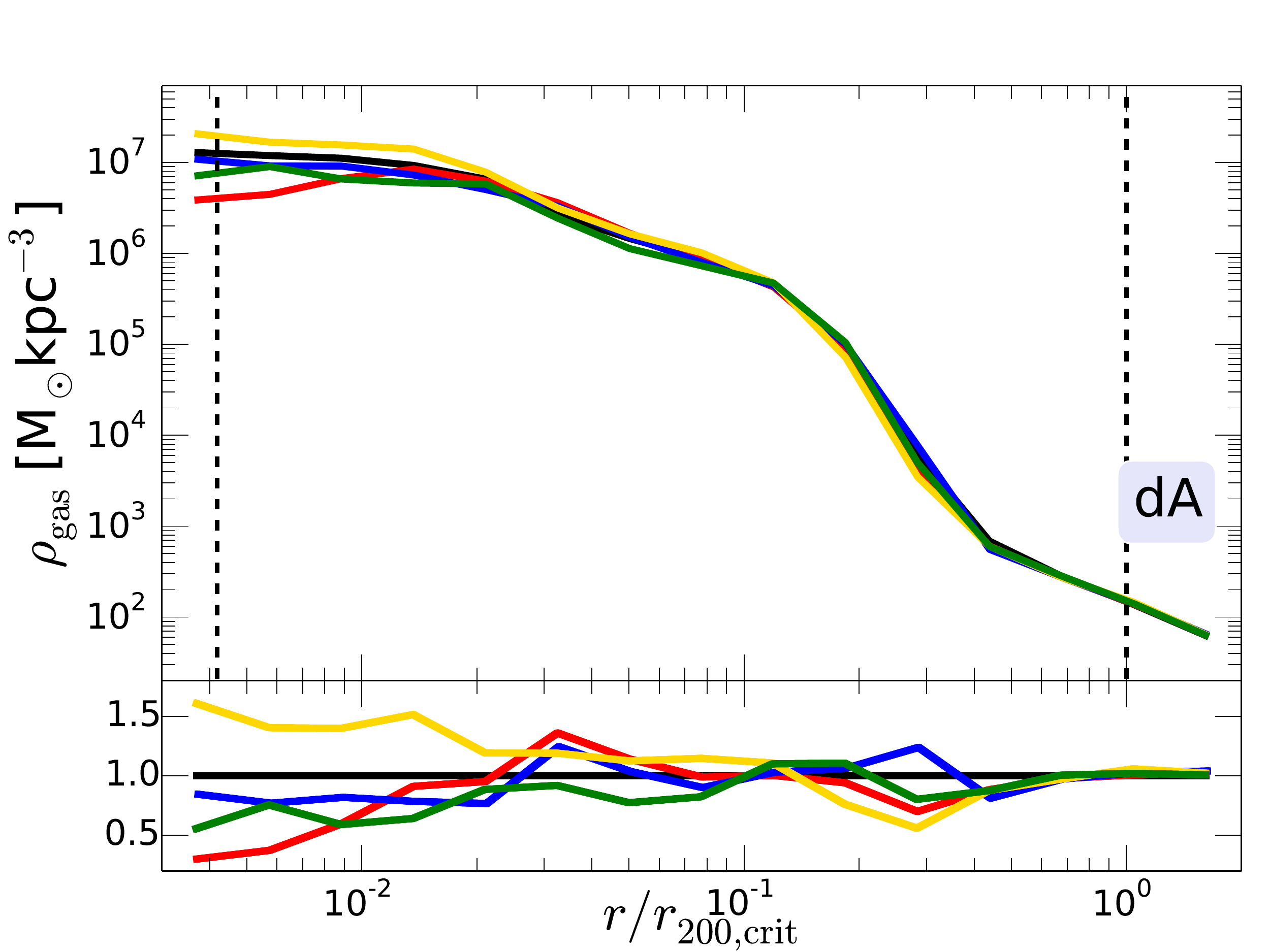}
\includegraphics[width=0.33\textwidth]{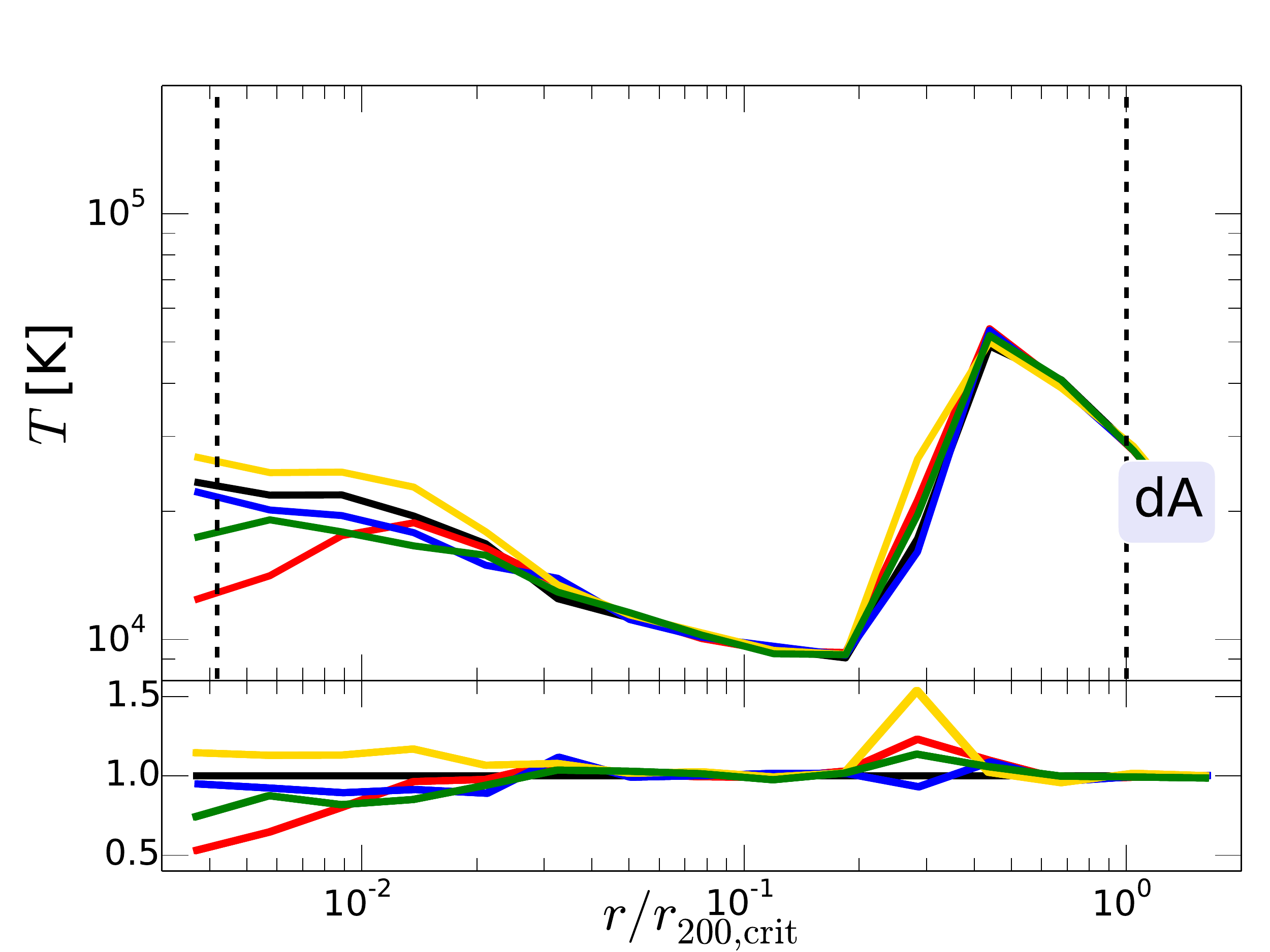}
\includegraphics[width=0.33\textwidth]{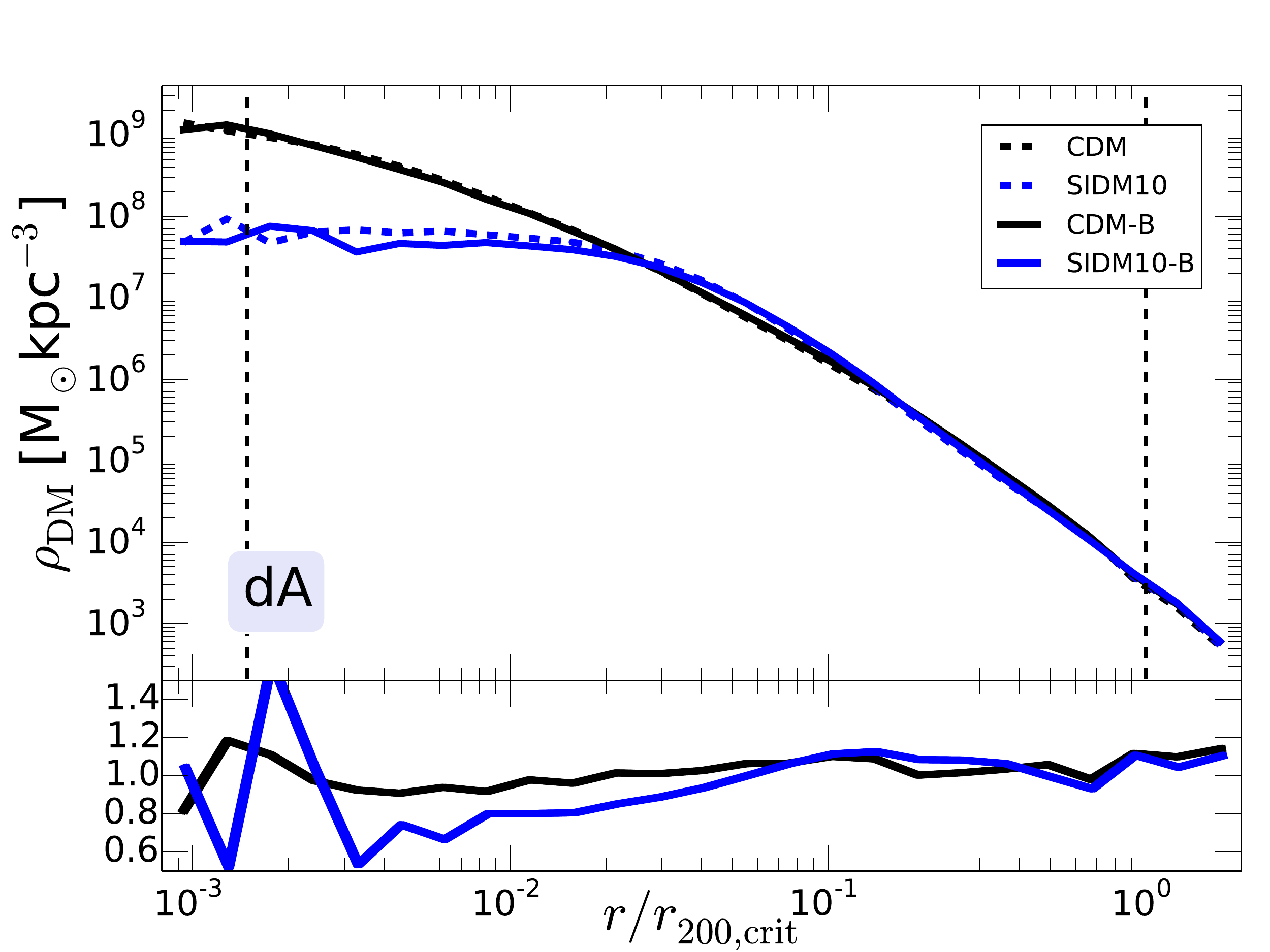}
\caption{Radial profiles for halo dA. We show from top left to bottom right: DM
density, DM velocity dispersion, DM velocity anisotropy, stellar density,
stellar velocity dispersion, stellar velocity anisotropy, gas density, gas
temperature, and in the last panel the DM density profile for the CDM and
SIDM10 models with and without baryons. Most profiles of halo dB look similar
with respect to the difference between the different DM models. However, the dB
halo is less relaxed due to its environment. This affects, for example, the
anisotropy ($\beta$) profile, which for dB  is not monotonically decreasing
towards the halo center. The dotted vertical lines mark $2.8$ times the
softening length and the mean virial radius.}
\label{fig:halo_profiles}
\end{figure*}

The bottom panels in Figure~\ref{fig:globalprop} show the relative differences
between the CDM case and the SIDM models. Most of these changes are small, of
the order of $\sim 10\%$, and not correlated with the specific DM model;~i.e.
there is no clear correlation with the cross section.  Nevertheless, there are
some interesting points. For instance, the total stellar mass seems to increase
in most of the SIDM models. The neutral hydrogen content on the other hand is
decreasing for most SIDM models compared to the CDM case. However, for both
observables the effect is at maximum around $10\%$.  The changes in the star
formation rate as a function of lookback time can be larger. Relative
differences in each time bin can be as large as $\sim 20-30\%$.  As for the
stellar mass, the largest differences occur for the SIDM10 model compared to
the CDM case. The high metallicity tail of the stellar MDF is also sensitive to
the DM model. However, this region of the MDF is not probed very well due to
low number statistics. The same is true for the low metallicity tail of the distribution.

We conclude that all models are in reasonable agreement within the
observational range although there might be potential discrepancies if the
simulated galaxies represent the median of the distribution of a larger
complete sample. We also conclude that allowed changes on the DM
self-scattering cross sections do not strongly affect the global properties of
the two dwarfs. Most changes are of the order of $10\%$ at maximum, and we do
not find any systematic trends with the specific DM model; i.e. these changes
are largely stochastic and not directly correlated with the magnitude and type
of SIDM cross section.

\section{The relevance of DM self-interactions}\label{interactions_sec}

It is expected that the impact of DM collisions will be most evident in the
central regions of the dwarfs where the average number of collisions per
particle across the entire history of the galaxy is larger than one. We will
therefore focus on this region from here on and only refer to our isolated
system dA.  In Figure~\ref{fig:halo_profiles} we show various radial profiles
at $z=0$. We show the DM density, velocity dispersion, and velocity anisotropy
profiles in the first row while the second shows the same quantities for the
stellar component. For the gas in the last row, we show the density and
temperature profiles only. The lower right panel of
Figure~\ref{fig:halo_profiles} compares the DM density profiles of the CDM and
SIDM10 model for a simulation with DM particles only (CDM, SIDM10) to the full
baryonic physics simulations (CDM-B, SIDM10-B). At the bottom of each panel we
show the relative differences with respect to the CDM case. 

The upper DM panels demonstrate that DM collisions generate an isothermal
density core with a flat velocity dispersion and a spatial extent that is
related to the magnitude of the scattering cross section at the typical DM
velocities of the central regions.  All the allowed SIDM models have core sizes
$\lesssim 2\kpc$ at the scale of dwarfs.  We also notice that while models with
a constant cross section predict a strong dependence of the core size with halo
mass~\citep{Rocha2012}, the Yukawa-like vdSIDM models we explore here naturally
create a much milder correlation. The behaviour of the velocity anisotropy
illustrates how DM collisions isotropize the orbits of the DM particles with an
amplitude that is correlated with the magnitude of the cross section: a larger
cross section leads to a lower anisotropy. 

One can clearly see that the central DM density is reduced by at least a factor
$2$ at the softening scale for all SIDM models. For the most extreme model,
SIDM10, the redistribution of DM particles also leads to a significant increase
($\sim 40\%$) of the DM density at about $\sim 6\%$ of $r_{\rm 200,crit}$.
This effect can also be seen for the allowed SIDM1 model although the excess is
much smaller in that case ($\sim 10\%$), and occurs at a slightly smaller
radius ($\sim 4\%$ of $r_{\rm 200,crit}$). The effects on the DM distribution
is always largest in the SIDM10 case. This is also true for the anisotropy
parameter $\beta_{\rm DM}$ which is essentially zero within $10\%$ of $r_{\rm
200,crit}$ for the SIDM10 model. The transition to this isotropic velocity
distribution is much smoother for the other SIDM models, but all of them reach
$\beta_{\rm DM} \sim 0$ in the inner part of the halo, whereas the CDM goes
down to slightly larger values only ($\beta_{\rm DM} \sim 0.1$).
Self-interactions play a role only in the inner part ($\lesssim1$~kpc) of the
halo such that the outer profiles agree well between the different models.
However, the most extreme model, SIDM10, shows significant deviations in the
density and velocity structure even out to $\sim 30\%$ of $r_{\rm 200,crit}$.

The stellar distribution is also clearly affected by self-interactions. The
relative differences between CDM and SIDM are largest for the inner density and
anisotropy profiles. The SIDM10, and SIDM1 models lead to a decrease in the
central stellar density of more than a factor of $2$. Although the SIDM10 model
is ruled out due to its large cross section, SIDM1 is still a possible CDM
alternative, which leads to a significant modifications of the central stellar
density. Similar to the DM density, the SIDM1 and SIDM10 stellar densities also
exceeds the stellar density of the CDM model around $5\%$ of $r_{\rm
200,crit}$. Interestingly, the velocity dispersion profile $\sigma_{\rm
stars}^{\rm 3D}$ is not altered significantly through self-interactions. Also
the velocity anisotropy profiles of the stars are more similar, between the
different simulations, than those of DM.

The gas density profile in the lowest row of the figure also shows deviations
of about a factor of two. Except for vdSIDMa, all models behave similar to the
findings of the DM and stellar profiles;~i.e. a significant reduction of the
gas density in the center. vdSIDMa, on the other hand, shows an increased gas
density at the center. Also, the central temperature of vdSIDMa is higher than
all other models.  For SIDM1, the temperature is only $50\%$ of the gas
temperature in the CDM gas. However, it seems that the changes in the gas are
less correlated with the actual cross section than those in the DM and stellar
component. For example, the largest differences in the gas density and
temperature can be seen for SIDM1 and not for the more extreme model SIDM10.
Also vdSIDMa shows the opposite behaviour compared to the other SIDM models.  

\begin{table*}
\begin{threeparttable}[b]
\begin{tabular}{lccccccccccccccc}
\hline
component  & $\rho$ profile                          & $\rho$ [${\rm M_\odot kpc^{-3}}$]  & $r_s$ [${\rm kpc}$] & $r_c$ [${\rm kpc}$]  & $\alpha$    & $\beta$ profile       & $A$ & $a$ & $\alpha_\beta$ & $b$ & $\beta_0$\\
\hline
CDM-B         \\
\hline
dark matter      & NFW Eq.~(\ref{NFW})                 & $4.41\times10^7$               & $1.97$              & --                 &  --      & Eq.~(\ref{beta_profile})   & $0.083$  &$1.265$   & $0.385$   &  $0.541$ & $0.071$    \\
stars (inner)    & cored/exp. Eq.~(\ref{rho_bar})    & $5.24\times10^6$               & --                & $0.46$               & $2.44$     & Eq.~(\ref{beta_profile})   & $42.29$  &$2.790$   & $5.082$   & $0.312$  & $-0.060$ \\
stars (outer)    & cored/exp. Eq.~(\ref{rho_bar})    & $1.05\times10^7$               & $0.80$              & --                 & --       & Eq.~(\ref{beta_profile})   & $42.29$  &$2.790$   & $5.082$   & $0.312$  & $-0.060$ \\
gas (inner)      & cored/exp. Eq.~(\ref{rho_bar})    & $1.16\times10^7$               & --                & $0.98$               & $3.00$     & Eq.~(\ref{beta_profile})   & --     & --     & --      & --   & -- \\
gas (outer)      & cored/exp. Eq.~(\ref{rho_bar})    & $6.98\times10^6$               & $1.88$              & --                 & --       & Eq.~(\ref{beta_profile})   & --     & --     & --      & --   & -- \\
\hline
SIDM1-B    \\
\hline
dark matter     & Burkert-like Eq.~(\ref{burkert})     & $3.34\times10^8$               & $1.00$              & $3.74$              & --        & Eq.~(\ref{beta_profile})   & $2.011$   & $2.382$    & $3.807$   & $0.262$  & $0.000$  \\
stars (inner)   & cored/exp. Eq.~(\ref{rho_bar})     & $6.44\times10^6$               & --                & $0.77$              & $2.49$      & Eq.~(\ref{beta_profile})   & $21.62$   & $3.222$    & $4.766$   & $0.358$  & $-0.017$ \\
stars (outer)   & cored/exp. Eq.~(\ref{rho_bar})     & $1.29\times10^7$               & $0.75$              & --                & --        & Eq.~(\ref{beta_profile})   &  $21.62$   & $3.222$    & $4.766$   & $0.358$  & $-$0.017 \\
gas\tnote{1}    & cored/exp. Eq.~(\ref{rho_bar})     & $7.43\times10^6$               & $1.80$              & --                & --        & Eq.~(\ref{beta_profile})   & --     & --     & --      & --   & -- \\
\hline
SIDM10-B    \\
\hline
dark matter      & Eq.~(\ref{SIDM10_rho})                    & $1.48\times10^8$               & --                & $1.55$        & $2.82$      & Eq.~(\ref{beta_profile})   & $0.727$  & $3.258$     & $4.358$    & $0.278$   &  $0.000$\\  
stars (inner)    & cored/exp. Eq.~(\ref{rho_bar})    & $7.62\times10^6$               & --                & $0.90$              & $2.37$      & Eq.~(\ref{beta_profile})   & $25.54$  & $3.475$     & $4.986$    & $0.368$  & $-0.049$\\
stars (outer)    & cored/exp. Eq.~(\ref{rho_bar})    & $1.40\times10^7$               & $0.74$              & --                & --        & Eq.~(\ref{beta_profile})   & $25.54$  & $3.475$     & $4.986$    & $0.368$  & $-0.049$\\
gas (inner)      & cored/exp. Eq.~(\ref{rho_bar})    & $9.01\times10^6$               & --                & $0.88$              & $1.69$      & Eq.~(\ref{beta_profile})   & --     & --     & --      & --   & -- \\
gas (outer)      & cored/exp. Eq.~(\ref{rho_bar})    & $5.56\times10^6$               & $2.10$              & --                & --        & Eq.~(\ref{beta_profile})   & --     & --     & --      & --   & -- \\
\hline
vdSIDMa-B    \\
\hline
dark matter     & Burkert-like Eq.~(\ref{burkert})     & $1.33\times10^9$               & $0.64$              & $5.13$              & --        & Eq.~(\ref{beta_profile})  & $1.032$   & $1.892$    & $2.672$   & $0.281$ & $0.000$\\
stars (inner)   & cored/exp. Eq.~(\ref{rho_bar})     & $5.82\times10^6$               & --                & $0.56$              & $2.80$      & Eq.~(\ref{beta_profile})  & $114.5$   & $2.816$    & $5.983$   & $0.289$ & $0.015$\\
stars (outer)   & cored/exp. Eq.~(\ref{rho_bar})     & $1.09\times10^7$               & $0.81$              & --                & --        & Eq.~(\ref{beta_profile})  & $114.5$   & $2.816$    & $5.983$   & $0.289$ & $0.015$\\
gas (inner)     & cored/exp. Eq.~(\ref{rho_bar})     & $1.23\times10^7$               & --                & $0.87$              & $2.91$      & Eq.~(\ref{beta_profile})  & --     & --     & --      & --   & -- \\
gas (outer)     & cored/exp. Eq.~(\ref{rho_bar})     & $9.47\times10^6$               & $1.68$              & --                & --        & Eq.~(\ref{beta_profile})  & --     & --     & --      & --   & -- \\
\hline
vdSIDMb-B    \\
\hline
dark matter     & Burkert-like Eq.~(\ref{burkert_2})   & $8.49\times10^7$               & $1.57$              & $0.30$              & --        & Eq.~(\ref{beta_profile}) & $0.142$    & $0.983$    & $0.286$  & $0.553$ & $0.000$\\
stars (inner)   & cored/exp. Eq.~(\ref{rho_bar})     & $5.66\times10^6$               & --                & $0.57$              & $2.66$      & Eq.~(\ref{beta_profile}) & $37.28$    & $3.424$    & $5.250$  & $0.361$ & $0.042$\\
stars (outer)   & cored/exp. Eq.~(\ref{rho_bar})     & $1.17\times10^7$               & $0.77$              & --                & --        & Eq.~(\ref{beta_profile}) & $37.28$    & $3.424$    & $5.250$  & $0.361$ & $0.042$\\
gas (inner)     & cored/exp. Eq.~(\ref{rho_bar})     & $1.82\times10^7$               & --                & $1.26$              & $3.90$      & Eq.~(\ref{beta_profile}) & --     & --     & --      & --   & -- \\
gas (outer)     & cored/exp. Eq.~(\ref{rho_bar})     & $6.23\times10^6$               & $1.90$              & --                & --        & Eq.~(\ref{beta_profile}) & --     & --     & --      & --   & -- \\
\hline
\hline
\end{tabular}
\caption{Best fit parameters to the DM, stellar and gas density and anisotropy
profiles. The different columns list: the name of the DM model and the
component under consideration, the profile that was fit to that component, and
the parameters of the profiles obtained for the best fit. In the case of the
double component fits for the stars and for the gas, the goodness of the fit
(Eq.~\ref{goodness}) is computed for the combined fit.  In the case of stars,
the fit is restricted to the region within $10\kpc$. On the left we list the best fit
parameters for the density profiles and on the right those for the velocity anisotropy
profiles. For the latter, we do not distinguish inner and outer regions for the
stars, and we do not give a profile for the gas. We stress that we provide
different DM density profiles for the different DM models since a single
parametric model cannot be used to achieve a good fit to all models.}
\label{table:fits}
\begin{tablenotes}
    \item[1] In this case the fit is poor in the inner regions ($\lesssim1.5\kpc$), and 
    thus, we use only the exponential gas profile instead of the two-component model as in the other cases.
  \end{tablenotes}
\end{threeparttable}
\end{table*}

The lower right panel of Figure~\ref{fig:halo_profiles} demonstrates that the
feedback associated with SNe does not alter the DM density distribution in our
model. This is not surprising since we do not employ a very bursty star
formation model, but a rather smooth star formation prescription. As a
consequence, the DM density profile is not affected at all by the formation of
the baryonic galaxy and the related feedback processes for the CDM case. The
SIDM models lead to core formation due to self-interactions of DM particles.
Such core makes it easier for SNe feedback to drive gas outwards, which should
cause some effect on the DM distribution. In fact, the lower right panel of
Figure~\ref{fig:halo_profiles} demonstrates that the DM density is slightly
reduced in the cored region even with a smooth feedback model like ours.
However, this effect is rather small and at maximum $\sim 40\%$ relative to the
SIDM10 simulation without baryons. This effect is therefore small compared to
the effect of self-interactions, which reduce the central DM density much more
significantly. 

So far we have discussed the relative differences between the different
profiles.  To quantify the spatial distribution of the DM and the baryons, gas
and stars, in more detail, we now find analytical fits to the spherically
averaged density distributions. We have found that the different DM
models require different density profiles profiles to achieve a reasonable
quality of the fits.

\begin{figure}
\centering
\includegraphics[width=0.49\textwidth]{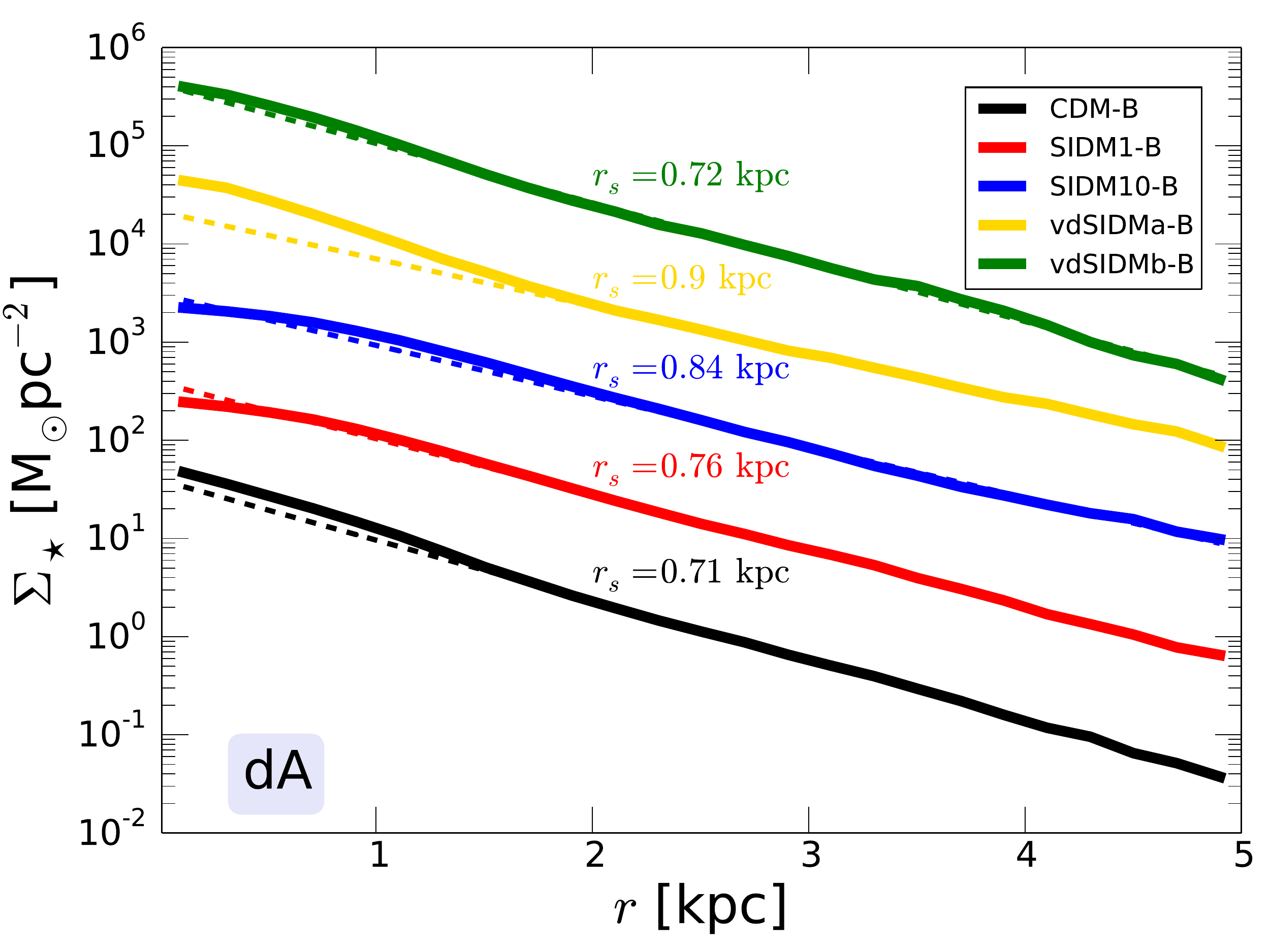}
\caption{Stellar surface density profiles with exponential fits (dashed lines)
for halo dA for all DM models. The non-CDM cases are shifted up by factors of
ten. Exponential scale lengths, $r_s$, are stated for each fit. The fits were
performed over the full radial range. All models lead to essentially perfect
exponential profiles with no significant bulge components. The largest
cross section models (SIDM1, SIDM10) produce a stellar core in the center.}
\label{fig:stellar_profiles}
\end{figure}

We start with the DM profile for the CDM case. It is well-known that CDM haloes
have spherically averaged density profiles that are well described by NFW
\citep[][]{Navarro1996} or Einasto profiles \citep{Springel2008}. We therefore fit the DM
profile of the CDM model with the two-parameter NFW profile:
\begin{equation}\label{NFW}
\rho_{\rm CDM}(r)=\rho_0\,\frac{r_s^3}{r(r+r_s)^2}.
\end{equation}
On the other hand, the SIDM haloes are well fitted by cored-like profiles that
vary according to the amplitude of the self-scattering cross section at the
typical velocities of the halo. In the case of the strongest cross section,
SIDM10, a good fit is obtained with the following three-parameter profile:
\begin{equation}\label{SIDM10_rho}
\rho_{\rm SIDM10}(r)=\rho_0\,\frac{r_c^\alpha}{(r_c^\alpha+r^\alpha)},
\end{equation}
while for intermediate cross sections, SIDM1 and vdSIDMa, a 
Burkert-like three-parameter formula provides a better fit:
\begin{equation}\label{burkert}
\rho_{\rm (SIDM1,vdSIDMa)}(r)=\rho_0\,\frac{r_s^3}{(r+r_c)\,(r^2+r_s^2)}.
\end{equation}
Finally, for the weakest cross section, vdSIDMb, a good fit is given by:
\begin{equation}\label{burkert_2}
\rho_{\rm (vdSIDMb)}(r)=\rho_0\,\frac{r_s^3}{(r+r_c)\,(r+r_s)^2}.
\end{equation}
Next we consider the profiles of the baryonic components.  For the stars and
the gas, we use a two component density profile: an exponential profile in the
outer region, which is a good approximation except for the gas beyond
$\sim20\kpc$, and a cored profile in the inner region, analogous to
Eq.~(\ref{SIDM10_rho}):
\begin{equation}\label{rho_bar}
 \rho_{\rm (\star,gas)}(r) = 
\begin{cases}
\rho_{\rm{out,(\star,gas)}} \, \exp(-\frac{r}{r_{s,{\rm(\star,gas)}}})   & r\geq r_{\rm in} \\
\rho_{\rm{in,(\star,gas)}}  \, \frac{r_c^\alpha}{r_c^\alpha+r^\alpha}              & r < r_{\rm in},
\end{cases}
\end{equation}
where we find that $r_{\rm in}=1.5\kpc$ provides a good fit in all cases except
for the gas distribution in the SIDM1 case.

For each profile (DM, gas, and stars), we find the best fit parameters by
minimising the following estimate of the goodness of the fit: 
\begin{equation}\label{goodness}
Q^2=\frac{1}{N_{\rm bins}}\sum_i\left({\rm ln} \rho_i(r_i) - {\rm ln} \rho_{\rm fit}(r_i)\right)^2,
\end{equation}
where the sum goes over all radial bins. We summarise the best fit parameters
for each component in Table~\ref{table:fits}.

We stress again that we need distinct parametric density profiles to better
describe the spatial DM structure of the halo for the different DM models.  For
instance, in the case of SIDM10, the value of $Q$ for the best fit using
Eq.~(\ref{SIDM10_rho}) is $0.004$, whereas using
Eqs.~(\ref{burkert}-\ref{burkert_2}) is $0.008$ and $0.074$, respectively. On
the other hand, for SIDM1, the values of $Q$ using
Eqs.~(\ref{SIDM10_rho}-\ref{burkert_2}), are, respectively:  0.020, 0.003,
0.021. Clearly, in this case, Eq.~(\ref{burkert}) is the best fit.

\begin{figure}
\centering
\includegraphics[width=0.49\textwidth]{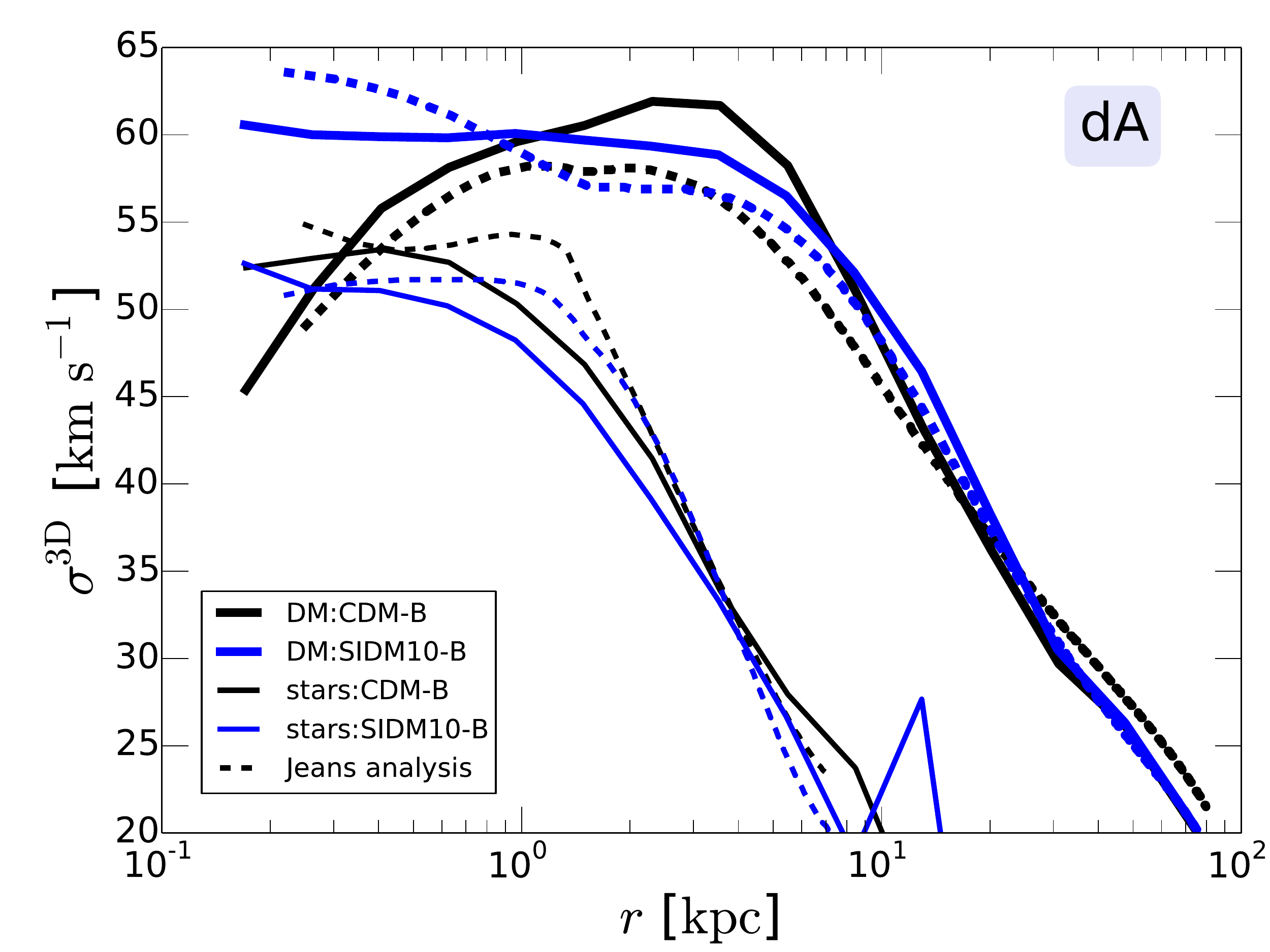}
\caption{Velocity dispersion profiles for CDM (black) and SIDM10 (blue)
compared to the results obtained with a Jeans analysis for halo dA. The DM
profile is shown with thick lines, whereas the stellar profile is shown with
thinner lines. Solid lines show the simulation results, whereas the results of
the Jeans analysis are shown with dashed lines.  The agreement between the
Jeans analysis and the simulation suggests that the galactic system (DM halo +
stars) is approximately in a collisionless spherical steady state. A similar
analysis cannot be performed for halo dB since this halo is not relaxed due its
merger history and environment, which is significantly more violent and less
isolated than that of dA.}
\label{fig:sigma_jeans}
\end{figure}

For the stars we can also inspect the stellar surface density profiles, which
are closely related to the measured stellar surface brightness profiles.  The
stellar surface density profiles of the dA dwarfs for the different DM models
are shown in Figure~\ref{fig:stellar_profiles}. The exponential scale length,
$r_s$, of the different models is quoted for each model, and the dashed lines
show the actual exponential fits for each model. For the CDM case, we find over
a large radial range an exponential profile and no significant bulge
contribution, similar to what is observed for most dwarfs. We have checked that
the surface density profiles do not vary much if the orientation of the galaxy
changes. The reason for this is that the dwarfs do not form thin disks, but
rather extended puffed up ellipsoidal distributions similar to, for example,
the stellar population of the isolated dwarf WLM.  The scale length values we
find are in reasonable agreement with other recent simulation of dwarf galaxies
at this mass scale \citep[e.g.,][]{Shen2013}. In the case of SIDM1 and SIDM10,
the presence of a small stellar core is visible in
Figure~\ref{fig:stellar_profiles}.  The scale length does not change
significantly as a function of the underlying DM model. However, it can clearly
be seen that DM self-interactions lead to slightly larger exponential scale
radii.

We note that, contrary to previous studies, we achieve exponential stellar
surface density profiles without a bursty star formation model
or a high density thresholds for star formation.  We therefore find that our
quiescent, smooth star formation model leads to non-exponential star formation
histories, and to exponential stellar surface density profiles.  It has been
argued that these characteristics are intimately connected to ``bursty'' star
formation rates~\citep[see e.g.][]{Governato2010}.  As a corollary, it was
argued that the formation of a DM core is then naturally expected. However, we
find that this is is not necessarily the case. 
We should note that \citet{Teyssier2013} simulated an
isolated dwarf of a similar halo mass and stellar mass as our dwarf dA but with a
considerably bursty star formation model that produced a $800\pc$ core.  This is in clear
contrast to our simulation where baryonic effects are unable to create a DM
core despite of the high global efficiency of star formation. The key is then,
once more, in the time scales and efficiency of energy injection during
SNe-driven outflows.  It remains to be seen if star formation histories in real
dwarf galaxies occur in bursts with a timescale much shorter than the local DM
dynamical timescale, and with an effective energy injection into the DM
particles that is sufficient to significantly alter the DM distribution.

As we have shown above, halo dA is in relative isolation and has a quiet
merger history. We therefore expect that the final stellar and DM configuration
is nearly in equilibrium. In the case of SIDM, once the isothermal core forms,
further collisions are not relevant anymore in changing the DM phase-space
distribution. We can then ignore the collisional term in the Boltzmann equation
and test the equilibrium hypothesis by solving the Jeans equation for the
radial velocity dispersion profile using as input the density and anisotropy
profiles:
\begin{equation}\label{Jeans_eq}
	\frac{1}{\rho}\frac{\rm d}{{\rm d}r}\left(\rho\sigma_r^2\right)+\frac{2\beta\sigma_r^2}{r}=-\frac{GM_{\rm tot}(<r)}{r^2},
\end{equation}
where $M_{\rm tot}(<r)$ is the total enclosed mass. We solve
Eq.~(\ref{Jeans_eq}) independently for the collisionless components, DM and
stars, using the fits to the density profiles with the analytic formulae
introduced above. In addition, we also fit the corresponding radial anisotropy
profiles for both the DM and the stars with the following five-parameter
formula:
\begin{equation}\label{beta_profile}
\beta(r)=A\left(\frac{r}{{\rm kpc}}\right)^a e^{-\alpha_\beta\left(\frac{r}{{\rm kpc}}\right)^b}+\beta_0,
\end{equation}
The best fit parameters for this relation for each DM model are listed in
Table~\ref{table:fits}.

The result obtained by solving the Jeans equation for the CDM and SIDM10 cases
is seen in Figure \ref{fig:sigma_jeans}. Here we show the predicted dispersion
profiles with dashed lines for DM (thick lines) and stars (thin lines). The
solid lines show the actual simulation results. Although the agreement between
the velocity dispersion predicted by the Jeans analysis and the simulation is
not perfect, the comparison still indicates that halo dA is roughly in
equilibrium and that the spherical approximations assumed above are partially
correct. In the SIDM10 case, this would suggest that the dark matter core
formed in the past and that any subsequent scattering does not affect the final
equilibrium configuration once the galaxy forms. This would justify  the use of
the Jeans equation without considering a collisional term. We will consider a
more detailed dynamical analysis in a subsequent paper analysing the different
SIDM cases, having a closer look at the velocity anisotropies, and also
investigating departures from spherical symmetry (Zavala \& Vogelsberger, in
prep).

\begin{figure}
\centering
\includegraphics[width=0.49\textwidth]{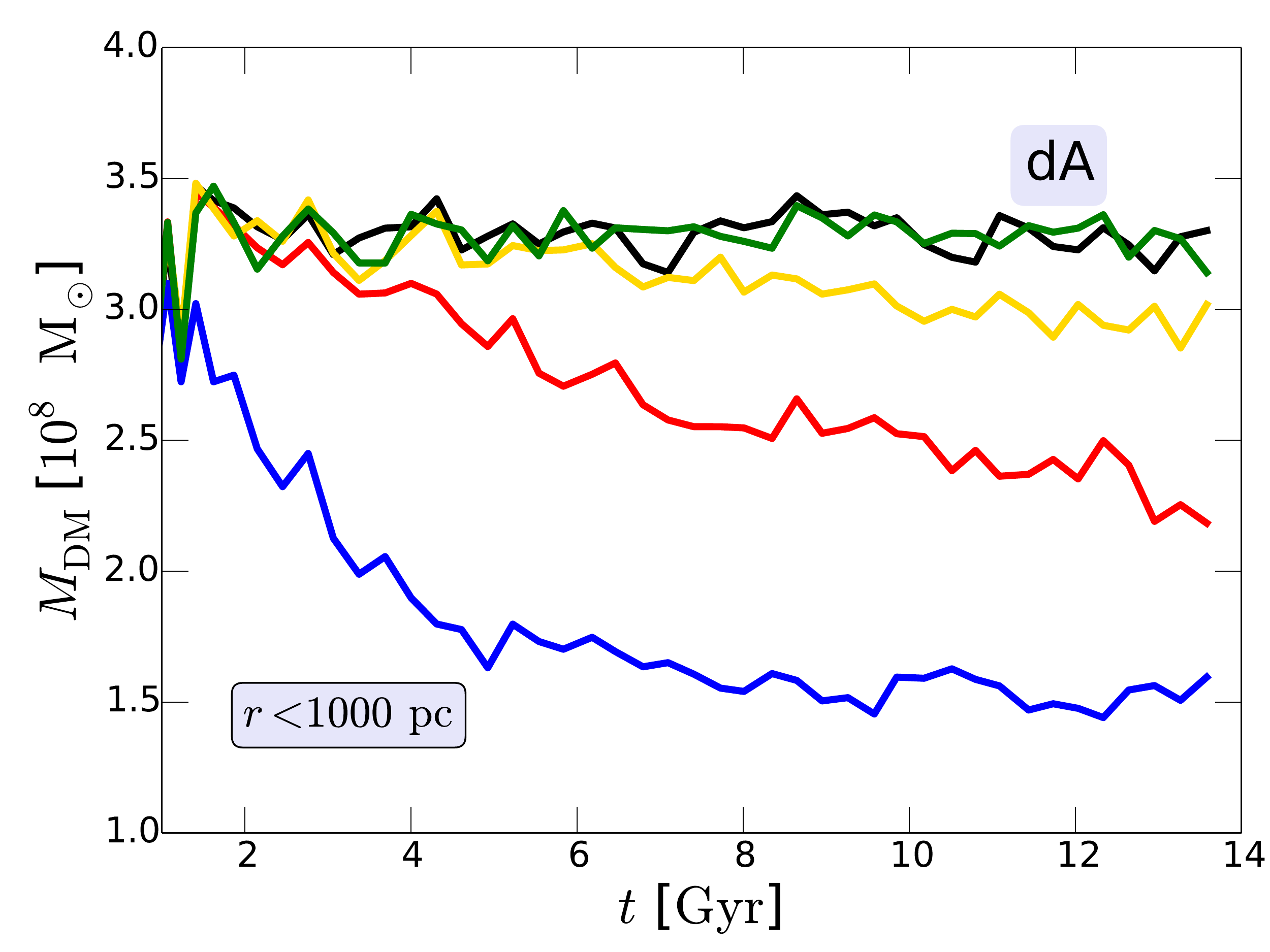}
\includegraphics[width=0.49\textwidth]{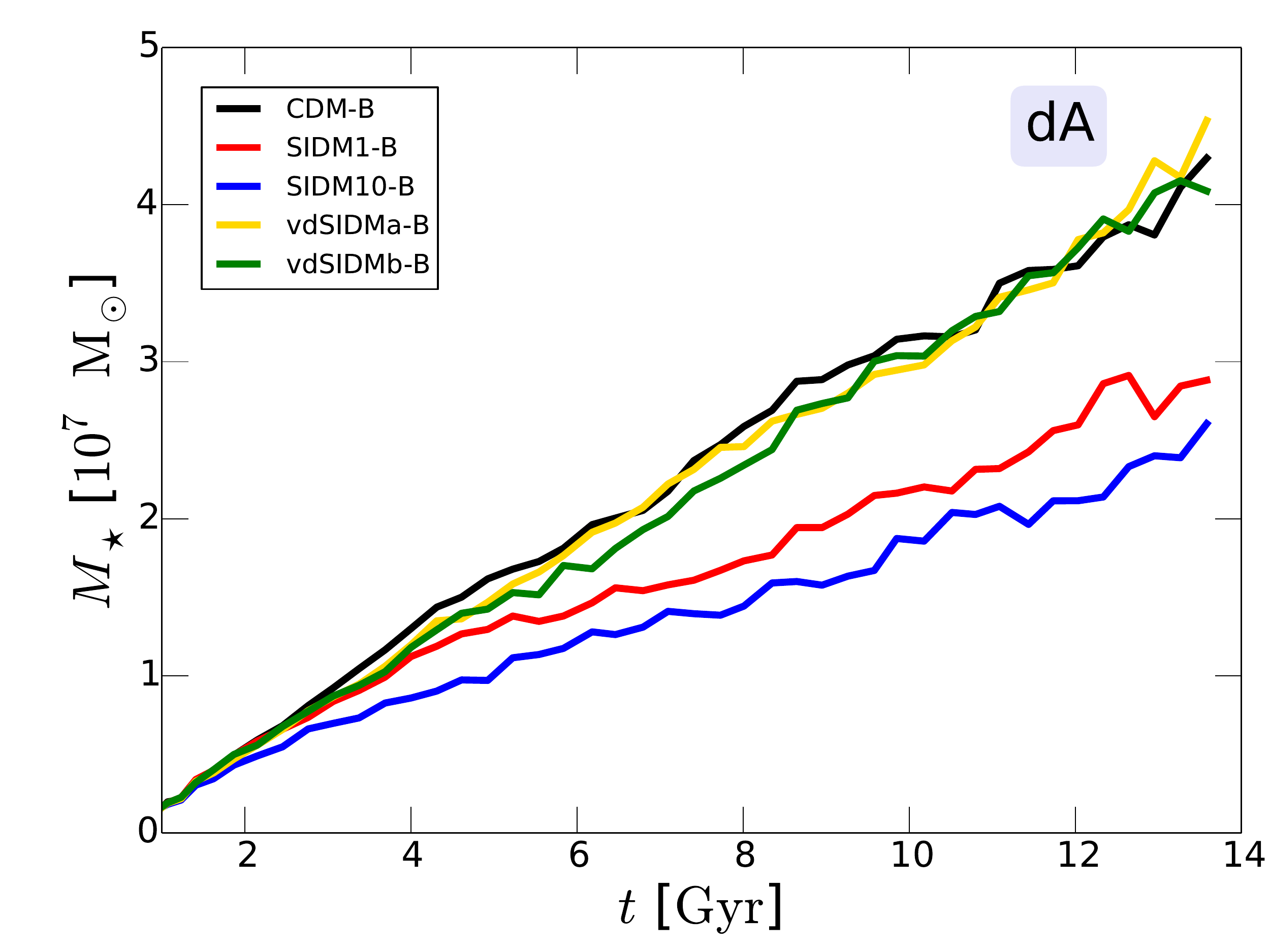}
\caption{Time evolution of the enclosed masses measured within $1000\pc$ for DM
(top) and stellar mass (bottom) for halo dA. The enclosed DM mass is for all
times and for all models significantly larger than the stellar mass, and
therefore dynamically dominates the center of the dwarf. The central DM mass is
substantially reduced for the SIDM1 and SIDM10 models, but only slightly for
the vdSIDM models. Similarly, the stellar mass is only reduced for the models
with constant cross section, whereas the stellar mass growth of vdSIDM closely
follows that of the CDM case.}
\label{fig:histories1000}
\end{figure}

\section{The inner halo}\label{sec_inner}

In this section we study in more detail the matter content and structure of the
simulated dwarf dA within the central region, $\sim1\kpc$, which roughly
encloses the DM core size for all models. 

We start with Figure~\ref{fig:histories1000}, which shows the mass buildup of
DM (top) and stars (bottom) within $1\kpc$ as a function of time.  In the cases
with a constant scattering cross section, it is clear that there is a
significant amount of dark matter mass expelled from the central kiloparsec. In
the case of SIDM1 for example, about $10^8\msun$ have been removed by $z=0$.
For the vdSIDM models however, there is only a minimal deviation from the
evolution of the base CDM model. In fact, the vdSIDMb model mass evolution
follows the CDM result very closely and shows a nearly constant central mass
after early times $\sim 1\Gyr$. The vdSIDMa model leads to a small depletion of
DM in the central $1\kpc$ of about $\sim 0.5\times 10^8\msun$. The largest
depletion can be seen for the SIDM10 model, where the central mass is reduced
by nearly a factor $3$.

The central stellar mass on the other hand grows steadily with time but it is
at all times, and for all DM models, sub-dominant compared to the inner DM
mass.  For all models the central stellar mass is below $5\times 10^7\msun$ at
$z=0$, which is a factor $\sim 5$ lower than the central DM mass at that time.
The stellar mass in SIDM1 and SIDM10 grows more slowly than in the CDM and vdSIDM
cases.  The vdSIDM models behave very similar to the CDM case, where the
stellar mass grows nearly linearly with time reaching a mass of about $4\times
10^7\msun$.  The stellar mass within $1\kpc$ grows initially similar SIDM10
(SIDM1),  however, after $\sim 2\Gyr$ ($\sim 4\Gyr$) the stellar mass growth
is slowed down for SIDM10 (SIDM1). After that time the growth is still linear
but with a significantly shallower slope compared to the CDM and vdSIDM cases.
We note that SIDM1 is an allowed model, and it is striking how different its
stellar mass is growing compared to the other allowed vdSIDM models.

\begin{figure}
\centering
\includegraphics[width=0.49\textwidth]{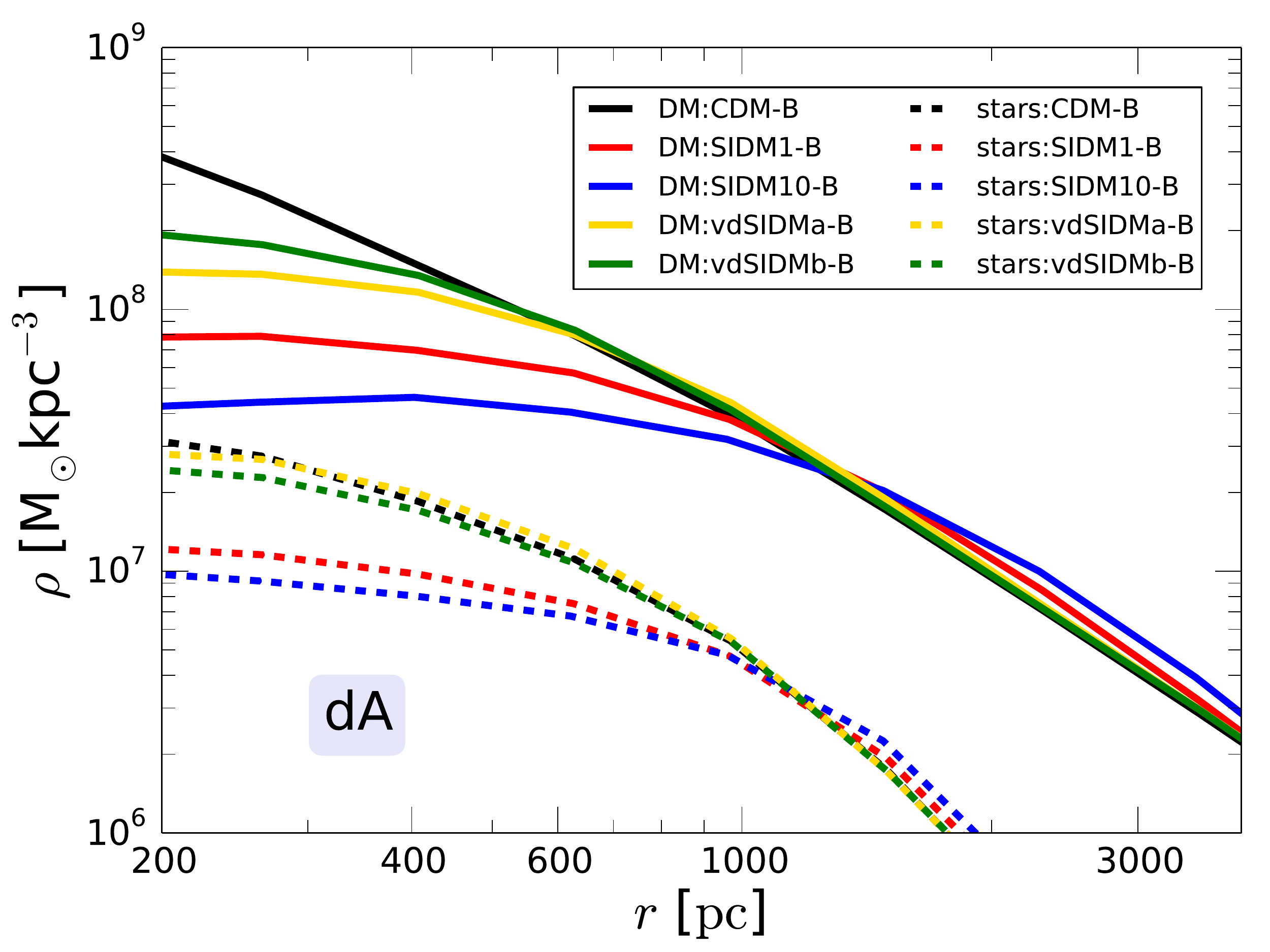}
\caption{Density profile of halo dA for DM (solid) and stars (dashed) within
the inner $4\kpc$ for the different DM models. The stars trace the evolution of
DM and also form a core. The size of the stellar core is closely related to the
size of the DM core. This can be seen most prominently for the SIDM1 and SIDM10
models.}
\label{fig:rho_inner}
\end{figure}

To quantify this in more detail we present a closer look of the density
profiles of DM (solid lines) and stars (dashed lines) in
Figure~\ref{fig:rho_inner}. This reveals a tight correlation between the shape
of the DM and stellar density distributions.  The stars within the core react
to the change in the potential of the dominant DM component due to
self-interactions. The size of the stellar core is therefore tied, to certain
degree, to the core sizes of the DM distribution. In the cases where the
scattering cross section has a velocity dependence, although the creation of a
DM core is evident, the impact is minimal in the stellar distribution compared
to the models with a constant cross section. This is mainly because even in the
CDM case, the stellar distribution forms a core which is roughly the size of
the DM core observed in the vdSIDM cases. We conclude that self-interactions
drive the sizes of the cores in DM and stars to track each other.  For SIDM1,
the density within the core is a factor of $\sim2-3$ smaller than in CDM. The
central distribution of stars can therefore probe the nature of DM and can
potentially be used to distinguish different SIDM models.

The strong correlation between DM and stars that we are finding is similar to
the one suggested recently by \cite{Kaplinghat2013} using analytical arguments,
but the regimes and interpretations are quite different. Whereas these authors
investigated the response of SIDM to a dominant stellar component, we are
investigating a system where DM still dominates dynamically. Thus in the
former, the DM cores sizes are reduced relative to expectations from DM-only
simulations due to the formation of the galaxy, while in the latter, the
stellar distribution of the galaxy responds to the formation of the SIDM core
by increasing its own stellar core relative to the CDM case. This regime is
therefore more promising to derive constraints for the nature of DM.

\begin{figure}
\centering
\includegraphics[width=0.49\textwidth]{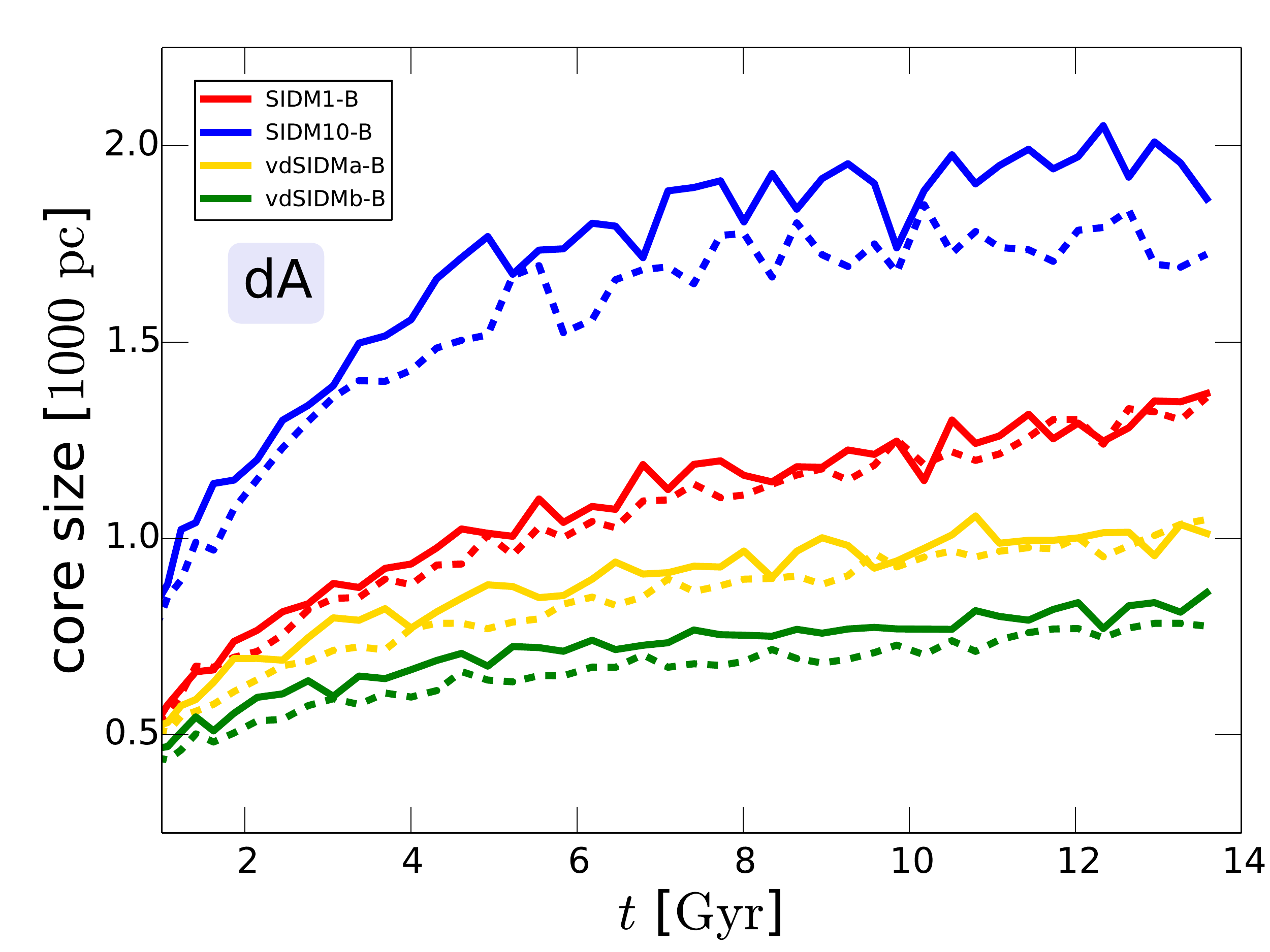}
\caption{DM core size as a function of time for halo dA. We compare the
evolution of the Burkert scale radius, $r_b$, in the DM-only simulations
(dashed) with the simulations including baryons (solid). Baryons have only a
tiny effect on the evolution and size of the cores. The largest effect can be
seen for SIDM10, where the shallow DM profile allows SNe feedback to expand the
core a bit more compared to the DM-only case.}
\label{fig:coresize}
\end{figure}

Next we are interested in the time evolution of the core radii.  It was already
obvious from Figure~\ref{fig:histories1000} that for the largest cross section
cases, the core should already be present early on during the formation history
of the galaxy. This is indeed the case as we demonstrate more clearly in
Figure~\ref{fig:coresize}, where the evolution of the core sizes are shown as a
function of time. As a measure of core radius, we fit Burkert
profiles~\citep[][]{Burkert1995} at each time, for each of the models, to
extract the core size $r_B$:
\begin{equation}\label{burkert_0}
\rho_B(r)=\rho_B\,\frac{r_B^3}{(r+r_B)\,(r^2+r_B^2)},
\end{equation}
We note that we use this two-parameter fit for simplicity to fit all SIDM
models and give a measure of the core size.  As we explored in detail above,
the different SIDM models are actually better fitted by different radial
profiles. However, our purpose here is not to rigorously define a core size but
simply to present an evolutionary trend for the different models. This trend is
clearly visible in the figure as well as the dependence of the amplitude of the
core size on the scattering cross section. Figure~\ref{fig:coresize} shows the
core radii determined by these two-parametric Burkert fits for all DM models
with (solid lines) and without (dashed lines) the effects baryons. 

Figure~\ref{fig:coresize} also demonstrates that the actual impact of baryons
on the DM distribution relative to the DM-only case is minor, as we discussed
already above (see lower right panel of Figure~\ref{fig:halo_profiles}).  In
the case of CDM this is not surprising since: (i) our star formation model is
less bursty compared to models where the cusp-core transformation is efficient
and (ii) for the mass scale we are considering, halo mass $\sim10^{10}\msun$
for halo dA, the energy released by SNe is not expected to be sufficient to
create sizeable DM cores
\citep{Governato2012,Penarrubia2012,Garrison-Kimmel2013}, although see
\citet{Teyssier2013}.  Figure~\ref{fig:coresize} demonstrates that our star
formation and feedback model creates only a slightly larger core for the SIDM10
model. This is because expelling gas in this case is easier due to the reduced
potential well caused by DM collisions. We stress again that these results are
sensitive to the model used for SNe-driven energy injection into the DM
particles (both efficiency and time scales). Larger efficiencies of energy
injection into shorter timescales would result in a larger removal of DM mass
from the inner halo. 

According to Figure~\ref{fig:coresize} a sizeable core is already
present very early on. By $t=4\Gyr$ all the models already have cores more than
half of their present day size.  Furthermore, Figure~\ref{fig:coresize}
also demonstrates, that none of our SIDM models lead to the gravothermal
catastrophe where the core collapses following the outward flux of energy
caused by collisions. This is consistent with the findings in VZL, where only
one subhalo, with similar total dark matter mass as halo dA, of the analogous
SIDM10 MW-size simulation was found to enter that regime towards $z=0$.

\begin{figure}
\centering
\includegraphics[width=0.49\textwidth]{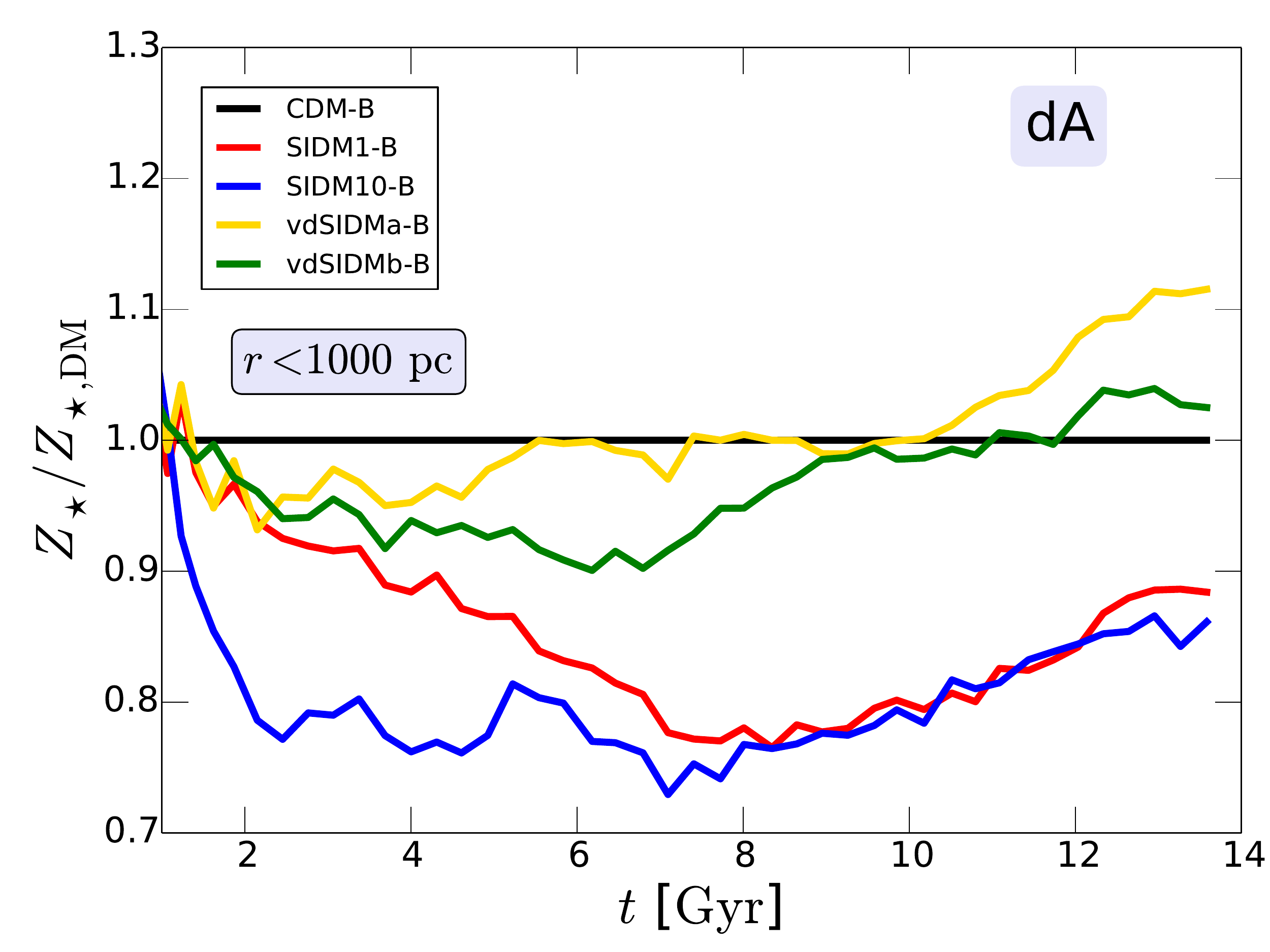}
\caption{Evolution of the central stellar metallicity (within 1 kpc). We show
the ratios of the different models with respect to the CDM case. The cases with
a constant cross section lead to a significant suppression of the central
stellar metallicity at $z=0$. The vdSIDM models have a weaker impact.}
\label{fig:metals}
\end{figure}

As a consequence of the DM core settling early on in the formation history of
the galaxy, the star formation rate within the central $1\kpc$ is reduced
significantly at late times in the cases with constant cross section. This
results in a stellar population that is in average older than in the case of
CDM. This is clearly shown in Figure~\ref{fig:metals}, where we plot the time
evolution of the ratio of the metallicity averaged within the central $1\kpc$,
relative to the CDM case. The difference today is $\gtrsim10\%$. Interestingly,
in the vdSIDM cases, there is an excess in star formation within $1\kpc$ in the
last stages of the evolution resulting in a younger stellar population since
the last $\sim2\Gyr$ (see also Figure~\ref{fig:coresize}). We will investigate
this issue, and in general the properties of the central $\sim\kpc$ region, in
a follow-up paper using simulations with increased resolution (Zavala \&
Vogelsberger, in prep).

\begin{figure}
\centering
\includegraphics[width=0.49\textwidth]{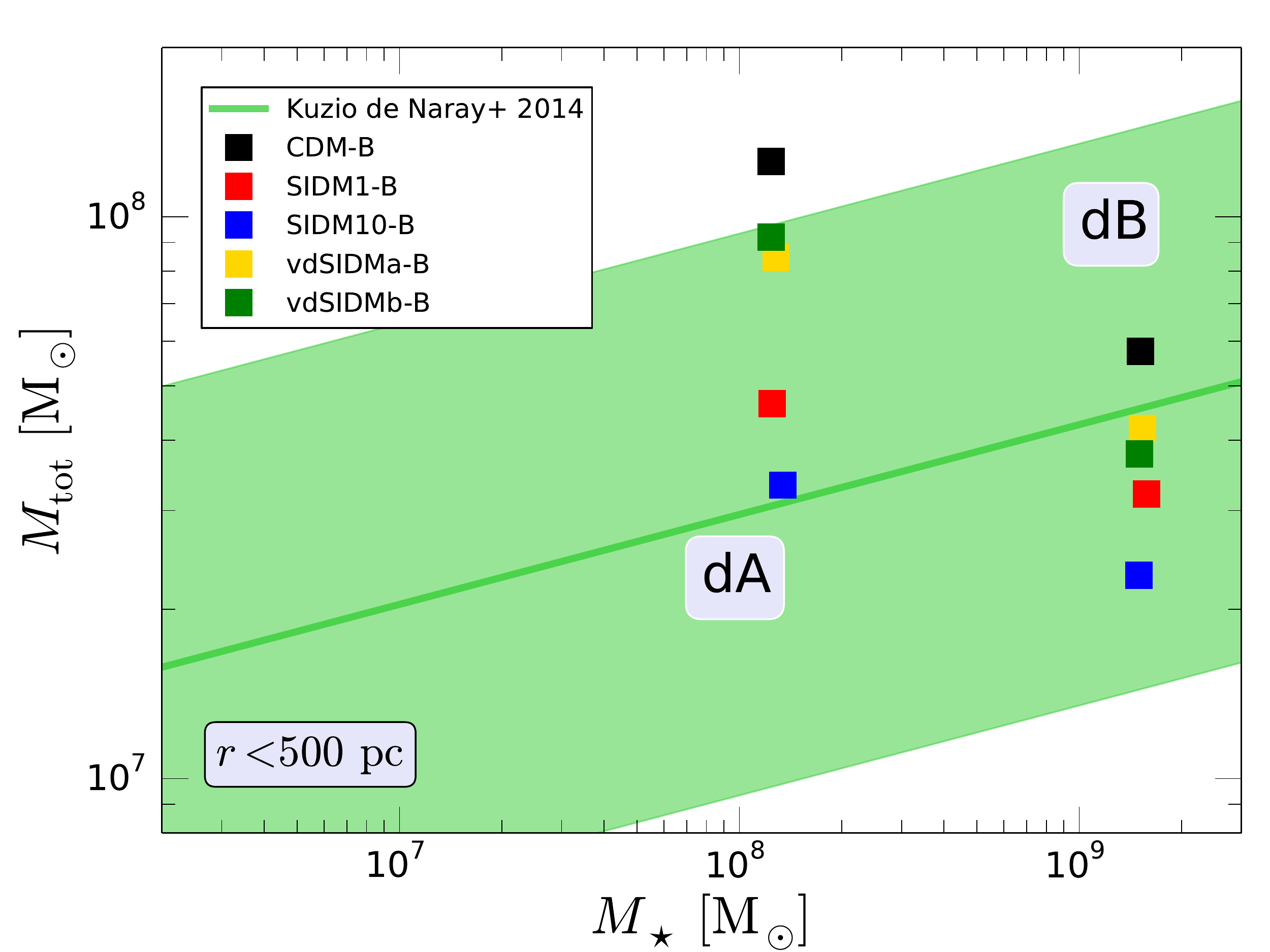}
\includegraphics[width=0.49\textwidth]{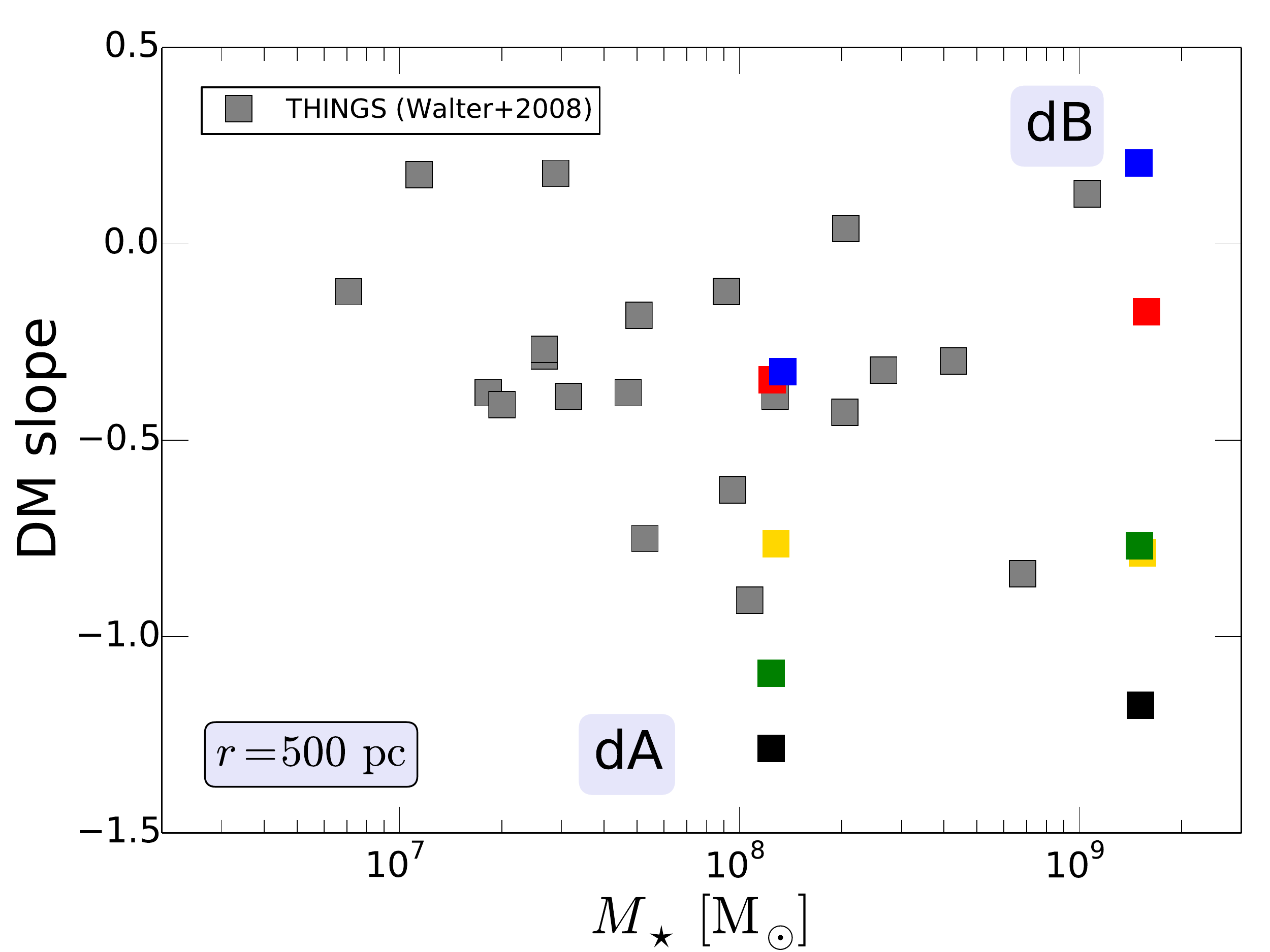}
\caption{Top panel: Enclosed total mass within $500\pc$ as a function of total
stellar mass. Bottom panel: DM density slope at $500\pc$ as a function of total
stellar mass. The different DM models lead to significantly different slopes
and masses at and within $500\pc$. At this radius even the vdSIDM models
clearly deviate from the CDM case. Both the mass and the slope clearly scale
with the cross section and allow to disentangle the different DM models.
Observational estimates from a combined sample of dwarf galaxies
\citep{KuziodeNaray2014} and from the THINGS survey \citep{Walter2008} are also
shown in the top and bottom panels, respectively.}
\label{fig:encmass500}
\end{figure}

In Figure~\ref{fig:encmass500} we focus on a region even closer to the halo
centre and show the total mass within $500\pc$ (top) and the slope of the
density profile measured at this radius (bottom). We compare both to
observational estimates using samples of dwarf galaxies
compiled in~\cite{KuziodeNaray2014} (top) and from the THINGS survey
\citep[bottom,][]{Walter2008}. At these small radii, the change in the enclosed
mass is still more dramatic for the constant cross section SIDM models having a
deficit in mass by a factor $\gtrsim3$ relative to the CDM case, while the
vdSIDM cases, although close to CDM, still deviate visibly.  The logarithmic
slope of the density profile at this radius varies between $-0.3$ (SIDM10) and
$-1.3$ (CDM).  Figure~\ref{fig:encmass500} shows that given the large
dispersion in the data, all DM models are essentially consistent with
observations. There is however some tension with the CDM simulation of halo dA
having a slightly too large total mass, and a slightly too steep DM density
slope at $r=500\pc$. On the other hand, the SIDM10 case might be to cored for
the stellar mass of halo dB ($M_\star\sim10^9\msun$). Taking both haloes into
account, and looking at the two relations of Figure~\ref{fig:encmass500} only,
it seems that SIDM1 agrees best with these observations. We stress however,
that our dwarf sample is far to small to draw any conclusions based on this
result and these observations are in any case, too uncertain to use them as
constraints.

\section{Conclusion}\label{concl_sec}

Self-Interacting Dark Matter (SIDM) is one the most viable alternatives to the
prevailing Cold Dark Matter (CDM) paradigm. Current limits on the elastic
scattering cross section between DM particles are set at
${\sigma/m_\chi<1\,{\rm cm}^2\,{\rm g}^{-1}}$~\citep{Peter2012}.  At this
level, the DM phase space distribution is altered significantly relative to CDM
in the centre of DM haloes. The impact of DM self-interactions on the baryonic
component of galaxies that form and evolve in SIDM haloes has not been explored
so far.  Recently, \cite{Kaplinghat2013} analytically estimated the DM
equilibrium configuration that results from a stellar distribution added to the
centre of a halo in the case of SIDM.  These authors studied the regime where
the stellar component dominates the gravitational potential and concluded that
the DM core sizes (densities) are smaller (higher) than observed in DM-only
SIDM simulations. This might have important consequences on current constraints
of SIDM models since they have been derived precisely in the baryon-dominated
regime.  In this paper we explore the opposite regime, that of dwarf galaxies
where DM dominates the gravitational potential even in the innermost regions.
Our analysis is based on the first hydrodynamical simulations performed in a
SIDM cosmology. We focus most of the analysis on a single dwarf with a halo
mass $\sim1.1\times10^{10}\msun$. We study two cases with a constant cross
section: SIDM1 and SIDM10, ${\sigma/m_\chi=1\, {\rm and}\, 10~{\rm cm}^2\,{\rm
g}^{-1}}$, respectively, and two cases with a velocity-dependent cross section:
vdSIDMa-b, that were also studied in detail in VZL and \cite{Zavala2013}.
Except for SIDM10, all these models are consistent with astrophysical
constraints, solve the ``too big to fail'' problem and create $\mathcal{O}$(1~kpc) cores in 
dwarf-scale haloes.

Our simulations include baryonic physics using the implementation described in
\cite{Vogelsberger2013} employing the moving mesh code {\tt AREPO}
\citep[][]{Springel2010}. We use the same model that was set up to reproduce
the properties of galaxies at slightly larger mass-scales. Our intention in
this first analysis is not to match the properties of dwarf galaxies precisely,
but rather to compare SIDM and CDM with a single prescription for the baryonic
physics, which has been thoroughly tested on larger scales. 

Our most important findings are:

\vspace{0.25cm}\noindent
{\bf Impact of SIDM on global baryonic properties of dwarf galaxies:} The
stellar and gas content of our simulated dwarfs agree reasonably well with
various observations including the stellar mass as a function of halo mass, the
luminosity metallicity relation, the neutral hydrogen content, and the
cumulative star formation histories. The latter are similar to those of local isolated group dwarf galaxies with
similar stellar masses.  We find that the stellar mass, the gas content, the
stellar metallicities and star formation rates are only minimally affected by
DM collisions in allowed SIDM models. The allowed elastic cross sections are
too small to have a significant global impact on these quantities, and the
relative differences between the different DM models are typically less than
$\sim 10\%$. In most cases these changes are not systematic as a function of
the employed DM model. The modifications in the global baryonic component of the
galaxies can therefore not be used to constrain SIDM models since the effects
are too small and not systematic. 

\vspace{0.25cm}\noindent
{\bf Impact of SIDM on the inner halo region:} Within
$\sim1\kpc$, we find substantial differences driven by the collisional nature
of SIDM. Besides the well-known effect of SIDM on the DM density profiles, we
also find that at these scales the distribution of baryons is
significantly affected by DM self-interactions. Both stars and gas show relative
differences up to $\sim 50\%$ in the density, the velocity dispersion, and the
gas temperature. Most of the effects increase with the size of the cross
section in the central region. The strongest correlation with the cross section
can be found for the stellar profiles, where the central stellar density
profile clearly correlates with the central cross section leading to lower
central densities for DM models with larger central cross sections.

\vspace{0.25cm}\noindent
{\bf Impact of baryons on the inner halo region:} We find that the impact of
baryons on the DM density profile is small for the DM-dominated dwarf
($M_\ast/M_{\rm DM}(<{\rm 1kpc})\lesssim0.15$) studied here. However, this
result is also connected to our smooth star formation model, which is not as
bursty as models where a significant core formation is observed due to baryonic
feedback. The size of the DM core and the central density are therefore
essentially the same as in our simulations that have no baryons, although the
core size is slightly larger in the former than in the latter. 

\vspace{0.25cm}\noindent
{\bf Disentangling different SIDM models:} For the cases where the scattering
cross section is constant, the combination of two key processes: (i) an early
DM core formation such that by $t=4\Gyr$, the DM cores already have half of
their size today; and (ii) a star formation history dominated by the period
after the formation of the DM core, result in the following characteristics of
the stellar distribution of SIDM galaxies: (a) The development of a central
stellar core with a size that correlates with the amplitude of the scattering
cross section. For instance, for the SIDM1 case with ${\sigma/m_\chi=1\,{\rm
cm}^2\,{\rm g}^{-1}}$, the density within the stellar core is a factor of
$\sim2-3$ smaller than for the CDM case. (b) A reduced stellar mass in the
sub-kpc region ($\gtrsim30\%$) as a byproduct of the reduced DM gravitational
potential due to self-scattering. (c) A reduced central stellar
metallicity; by $\gtrsim10\%$ at $z=0$ compared to the CDM
case. Around $z\sim 1$ the metallicity can be reduced by up to $\sim
25\%$.

For the cases where the scattering cross section is velocity-dependent, even
though a sizeable DM core can still be created ($\sim400\pc$), the effect in
the stellar distribution at all scales is minimal relative to CDM. This is
likely because the amplitude of the cross section within the inner region of
the dwarf is not large enough to produce a DM core that is larger than the
stellar core that forms in the CDM case. Whether the latter could be the result
of numerical resolution is something we will investigate in a forthcoming
paper. Any changes that we found in the vdSIDM cases seem to be only related to
the stochastic nature of the simulated star formation and galactic wind
processes.

These conclusions are key predictions of SIDM that can in principle be tested
to either constrain currently allowed models, particularly constant cross
section models, or to find signatures of DM collisions in the properties of the
central stellar distributions of dwarf galaxies. In future works we will
explore these possibilities in more detail.

\section*{Acknowledgements}

We thank Daniel Weisz for providing cumulative star formation histories of
local group dwarfs to us, and Michael Boylan-Kolchin for help with the initial
conditions. We further thank Volker Springel for useful comments and giving us
access to the {\tt AREPO} code. The Dark Cosmology Centre is funded by the
DNRF. JZ is supported by the EU under a Marie Curie International Incoming
Fellowship, contract PIIF-GA-2013-627723. The initial conditions were made
using the DiRAC Data Centric system at Durham University, operated by the
Institute for Computational Cosmology on behalf of the STFC DiRAC HPC Facility
(www.dirac.ac.uk). The DiRAC system is funded by BIS National E-infrastructure
capital grant ST/K00042X/1, STFC capital grant ST/H008519/1, STFC DiRAC
Operations grant ST/K003267/1, and Durham University. DiRAC is part of the Re:
green card UK National E-Infrastructure.

\label{lastpage}

\end{document}